\documentclass[english,11pt,nofootinbib,showpacs,notitlepage]{revtex4-1}
\usepackage[T1]{fontenc}
\usepackage{babel}
\usepackage[utf8]{inputenc}
\usepackage{amssymb,mathrsfs,amsfonts,amsmath,euscript,textcomp,graphicx,wrapfig,color,txfonts,lmodern,multirow,bigdelim,dsfont}
\usepackage[section]{placeins}
\usepackage[usenames,dvipsnames,svgnames,table]{xcolor}
\definecolor{darkgreen}{rgb}{0,.7,0}
\usepackage[colorlinks,linkcolor=blue,citecolor=darkgreen,pagebackref=true]{hyperref}
\usepackage{float}
\def\dac{\displaystyle\frac}

\def\[{\left[}
\def\]{\right]}
\def\({\left(}
\def\){\right)}

\def\1{{\bf CI}}
\def\2{{\bf CII}}
\def\3{{\bf CIII}}
\def\i{\mathcal{I}}
\def\j{\mathcal{J}}
\def\x{\mathcal{X}}
\def\y{\mathcal{Y}}
\def\c{\mathcal{C}^1}
\def\cc{\mathcal{C}^2}
\def\E{\mathcal{E}}

\def\z{\mathbb{Z}_{\geq0}}
\def\Lall{$L_1+L_2+L_3~$}

\newcommand{\vsep}{\hspace{.5cm}\vrule\hspace{.5cm}}
\newcommand{\hsep}{\hrule\vspace{.3cm}\nd}
\newcommand{\eq}[1]{\begin{equation}#1\end{equation}}
\newcommand{\eqs}[1]{\begin{equation*}#1\end{equation*}}

\newcommand{\const}{\mathop{\rm const}\nolimits}

\newcommand{\e}{\mathop{\rm e}\nolimits}

\newcommand{\br}[1]{\left\{#1\right\}}

\newcommand{\bgbr}[1]{\bigl\{#1\bigr\}}

\newcommand{\bgsb}[1]{\bigl[#1\bigr]}
\newcommand{\Bgp}[1]{\Bigl(#1\Bigr)}
\newcommand{\bgp}[1]{\bigl(#1\bigr)}

\newcommand{\nd}{\noindent}

\newenvironment{tightcenter}{%
  \setlength\topsep{0pt}
  \setlength\parskip{0pt}
  \begin{center}
}{%
  \end{center}
}

\begin{document}

\baselineskip7mm
\title{Non-constant volume exponential solutions in higher-dimensional Lovelock cosmologies}

\author{Dmitry Chirkov}
\affiliation{Sternberg Astronomical Institute, Moscow State University, Moscow 119991 Russia}
\affiliation{Faculty of Physics, Moscow State University, Moscow 119991 Russia}
\author{Sergey A. Pavluchenko}
\affiliation{Instituto de Ciencias F\'isicas y Matem\'aticas, Universidad Austral de Chile, Valdivia, Chile}
\author{Alexey Toporensky}
\affiliation{Sternberg Astronomical Institute, Moscow State University, Moscow 119991 Russia}
\affiliation{Kazan Federal University, Kazan 420008, Russia}

\begin{abstract}

In this paper we propose a scheme which allows one to find all possible exponential solutions of special class -- non-constant volume solutions -- in
Lovelock gravity in arbitrary number of dimensions and with arbitrate combinations of Lovelock terms.
We apply this scheme to (6+1)- and (7+1)-dimensional flat anisotropic cosmologies in Einstein-Gauss-Bonnet and third-order Lovelock gravity to
demonstrate how our scheme does work. In course of this
demonstration we derive all possible solutions in (6+1) and (7+1) dimensions and compare solutions and their abundance between cases with different
Lovelock terms present. As a special but more
``physical'' case we consider spaces which allow three-dimensional isotropic subspace for they could be viewed as examples of compactification schemes. Our results suggest that the same solution with
three-dimensional isotropic subspace is more ``probable'' to occur in the model with most possible Lovelock terms taken into account, which could be used as kind of anthropic argument for 
consideration of Lovelock and other higher-order gravity models in multidimensional cosmologies.
\end{abstract}

\pacs{04.20.Jb, 04.50.-h, 98.80.-k}


\maketitle

\section{Introduction}

Multidimensional paradigm being rather popular and motivated  by a number of different
approaches (from string theory to anthropic principle based on very specific and useful
for life properties of gravitation in 3 dimensions) have also well known difficulties.
We need to hide extra dimensions, so that it is reasonable to expect that they are
contracting (or at least have been contracting during some period of Universe evolution),
from the other hand three large dimensions are expanding almost isotropically. Such combination
of expanding and contracting dimensions can be achieved in theories with higher order
curvature corrections, however, the nature of this qualitative difference between large and small dimensions
is often completely obscure. Many papers even start from decomposition of metrics as a product
of ``external'' isotropic and ``internal'' spaces thus postulating this difference without any
attempt to shed light on its origin.

That is why any situation in which this division is not postulated {\it a priori}, but appeared
due to some underlying principle is of particular interest. One of such situation have been
found recently in Gauss-Bonnet gravity. Initially the solutions in question appeared when
generalisation of Kasner solution for a flat anisotropic Universe have been studied in the
regime when the Gauss-Bonnet term is dominated . It was found that apart from a solution in which
scale factor have a power-law behavior \cite{Deruelle1, Deruelle2} (a direct analog of the Kasner solution) a solution with
exponential time dependence of the scale factor exists \cite{Ivashchuk}. It has no analogue in General Relativity.
Later this class
solution have been generalized to  full Einstein-Gauss-Bonnet (EGB) theory (where we do not neglect
the Einstein term in the action) as well as to the theory with a cosmological constant \cite{grg10, KPT}.
In a more general setup such solutions (dubbed anisotropic inflation) have been shown to exist
in theories with Ricci square corrections \cite{Barrow2, Barrow1} , though being absent in $f(R)$ gravity, where only
Gauss curvature enters into the action.

An analysis of existence of such solutions in EGB gravity reveals an interesting fact that
they exists only if the space has isotropic subspaces~\cite{CST1} (though not necessary 3-dim
isotropic subspaces). The only exception is so-called constant-volume solutions~\cite{CST2}, but
they are special class and we do not consider them in this paper.
Though applicability of these solutions to a realistic description
of our Universe still needs more efforts in order to incorporate matter and exit from early
time inflation, the fact that isotropic subspaces appears as a condition for this solution
to exist is rather promising.

One of the first attempts to find an exact static solutions with metric being a cross product of a (3+1)-dimensional manifold times a constant curvature
``inner space'', also known as ``spontaneous
compactification'', were done in~\cite{add_1}, but with four dimensional Lorentzian factor being actually Minkowski (the generalization for a constant
curvature Lorentzian manifold was done
in~\cite{Deruelle2}). In the cosmological context it could be useful to consider Friedman-Robertson-Walker as a manifold for (3+1) section; this situation
with constant sized extra dimensions was
considered in~\cite{add_4}. There it was explicitly demonstrated that to have more realistic model one needs
to consider the dynamical evolution of the extra dimensional scale factor as well. In the context of exact solutions such an attempt was done
in~\cite{Is86}
where both the (3+1) and the extra dimensional scale factors where exponential functions. Solutions with exponentially increasing (3+1)-dimensional scale
factor and exponentially shrinking extra dimensional scale factor were described.

Of recent attempts to build a successful compactification particularly relevant are \cite{add13} where
the dynamical compactification of (5+1) EGB model was considered, \cite{MO04, MO14}, with different metric {\it ansatz} for scale factors corresponding to
(3+1)- and extra dimensional parts,
and \cite{CGP1, CGP2} where general (e.g. without any {\it ansatz}) scale factors and curved manifolds were considered.

In~\cite{Deruelle2} the structure of the equations of motion for
Lovelock theories for various types of solutions has been studied. It was stressed that the Lambda term in the action is actually not a
cosmological constant as it does not give the curvature scale of a maximally symmetric manifold. In the same paper the equations of motion for
compactification with both time dependent scale factors were written for arbitrary Lovelock order in the special case that both factors are flat.
The results of~\cite{Deruelle2} were reanalyzed for the special case of 10 space-time dimensions in~\cite{add_10}.
In~\cite{add_8} the existence of dynamical compactification solutions was studied with the use of Hamiltonian formalism.

Usually when dealing with cosmological solutions in EGB or more general Lovelock gravity~\cite{Lovelock} one imposes a certain {\it ansatz} on the metric.
Two most used (and so well-studied)
are power-law and exponential {\it ansatz}. The former of them could be linked to Friedman (or Kasner) stage while the latter -- to inflation. Power-law
solutions were intensively studied
some time ago~\cite{Deruelle1, Deruelle2} and recently~\cite{mpla09, prd09, Ivashchuk, prd10, grg10} which leads to almost complete their description (see
also~\cite{PT} for useful comments
regarding physical branches of the solutions). Exponential solutions, on the other hand, for some reason are less studied but due to their
``exponentiality'' could compactify extra dimensions
much faster and more reliably. Our first study of exponential solutions~\cite{KPT} demonstrate their potential and so we studied exponential solutions in
EGB gravity full-scale. We described
models with both variable~\cite{CST1} and constant~\cite{CST2} volume and developed general solution-building scheme for EGB. And now we are generalizing
this scheme for general Lovelock gravity.

In the present paper we describe a general scheme which allows us to get such solution
in an arbitrary Lovelock gravity with arbitrary number of dimensions. This scheme is illustrated
by third order Lovelock theory in (6+1) and (7+1) dimensions.

As one of the objectives we want to compare solutions in the same dimensionality but with different Lovelock terms taken into consideration. To be
specific, we will compare
(6+1)-dimensional solutions in EGB with (6+1)-dimensional solutions in $L_1 + L_2 +L_3$; then the same will be performed with (7+1)-dimensional solutions.
The reasoning behind this
comparison is simple -- originally, it was EGB gravity which was motivated by the string theory, and third-order curvature correction which comes from
string theory consideration,
do not coincide with third Lovelock term and, through that, are not ghost-free (see~\cite{zw, zum}). With this in mind general Lovelock theory is
worse-motivated then EGB, but in
higher dimensions, if we want to ``estimate'' the influence of higher-order terms, Lovelock theory could give us some ``insight''. Yet, if we want to
formally follow what comes
from M/string theory, we need to stay with EGB -- with both these arguments at hand, we will consider them both and compare the solutions one could get
from both of them.

Another important and more ``physical'' task is to explore the abundance of the solutions with three-dimensional isotropic subspace. Indeed, when dealing
with higher-dimensional cosmological models,
one needs to keep in mind that we observe only three spatial dimensions. This way we pay special attention to spatial splitting which have
three-dimensional isotropic subspace -- if these three
dimensions expand while the remaining directions contract, this would be successful dynamical compactification scheme in action.

The structure of the manuscript is as follows: first we introduce the most general equations we are dealing with. Then we develop the scheme for finding
all possible spatial splittings in any number
of dimensions and with any possible Lovelock terms and their combinations. After that in Section~\ref{application} we apply the scheme for (7+1)-dimensional space-times for \Lall case to
retrieve all possible spatial splittings. Then in Section~\ref{solutions} we describe all possible splittings in (6+1) and (7+1) dimensions and the
corresponding solutions. After that we separate solutions which allow three-dimensional isotropic
subspaces. Finally we draw conclusions and discuss obtained results.

\section{The Set-up.\label{setup}}
Let us spell out the conventions that we will use throughout this work. We choose to use units such that speed of light and gravitational constant are
equal to 1; Greek indices run from 0 to D, while Latin one from 1 to D unless otherwise stated; we also will use Einstein summation convention.

Let us consider $(D+1)$-dimensional flat space-time with Lovelock gravity. The gravitational action is
\eq{S=\frac{1}{2\kappa^2}\int d^{D+1}x\sqrt{|g|}\bgbr{\mathcal{L}+\mathcal{L}_m},\quad\mathcal{L}=\sum\limits_{n=0}^{d}c_n\mathcal{R}_n,\quad
d=\left\lfloor\frac{D}{2}\right\rfloor,\quad\mathcal{R}_n=\frac{1}{2^n}\Delta_{\alpha_1\ldots\alpha_{2n}}^{\beta_1\ldots\beta_{2n}}\prod\limits_{s=1}^{n}
R^{\alpha_{2s-1}\alpha_{2s}}_{\beta_{2s-1}\beta_{2s}}\label{action}}
where $\kappa^2$ is the $(D+1)$-dimensional gravitational constant, $\mathcal{L}_m$ is the Lagrangian of a matter, $g$ is the determinant of the metric
tensor, $R^{\alpha\beta}_{\mu\nu}$ stands for the components of the Riemann tensor, $c_n$ are constants, $\left\lfloor\frac{D}{2}\right\rfloor$
corresponds
to the integer part of $\frac{D}{2}$; the generalized Kronecker delta is defined as:
\eq{\Delta_{\alpha_1\ldots\alpha_{2n}}^{\beta_1\ldots\beta_{2n}}=\det
\left|
            \begin{array}{ccc}
              \delta^{\beta_1}_{\alpha_1} & \ldots & \delta^{\beta_{2n}}_{\alpha_1} \\
              \vdots & \ddots & \vdots \\
              \delta^{\beta_1}_{\alpha_{2n}} & \ldots & \delta^{\beta_{2n}}_{\alpha_{2n}} \\
            \end{array}
\right|}
We choose a reference system in such a way that metric has the following form:
\eq{ds^2=-dt^2+\sum\limits_{k}\e^{2a_k(t)}dx_k^2\label{g}}
Hereafter we will write $a_k,\dot{a}_k,\ddot{a}_k$ instead of $a_k(t),\dot{a}_k(t),\ddot{a}_k(t)$ for brevity. It is easily shown that
\eq{R^{0i}_{0i}=\ddot{a}_i+\dot{a}_i^2,\quad R^{j_1j_2}_{j_1j_2}=\dot{a}_{j_1}\dot{a}_{j_2},\;\;j_1<j_2,\quad
R^{\alpha\beta}_{\mu\nu}=0,\;\;\bgbr{\alpha,\beta}\ne\bgbr{\mu,\nu},\label{components.of.the.Riemann.tensor}}
the dot denotes derivative w.r.t. $t$. So, arbitrary component of the Riemann tensor takes the form:
\eq{R^{\mu\nu}_{\lambda\sigma}=
\left\{\sum\limits_{k}\left(\ddot{a}_k+\dot{a}_k^2\right)\delta_0^{[\mu}\delta_k^{\nu]}\delta^0_{[\lambda}\delta^k_{\sigma]}+
\sum\limits_{i<j}\dot{a}_i\dot{a}_j\delta_i^{[\mu}\delta_j^{\nu]}\delta^i_{[\lambda}\delta^j_{\sigma]}\right\},\label{gen.expr.for.Riemann}}
square brackets denote the antisymmetric part on the indicated indices. It can be shown that
\eqs{\frac{\sqrt{|g|}\mathcal{R}_n}{2^n}=\left\{\mathfrak{S}_{n}\sum\limits_{j_1<\ldots<j_{2n}}\prod\limits_{r=1}^{2n}\dot{a}_{j_r}+
n\mathfrak{S}_{n-1}\left[\sum\limits_{k}\dot{a}_{k}^2
\sum\limits_{\{j_1<\ldots<j_{2n-2}\}\ne k}\prod\limits_{r=1}^{2n-2}\dot{a}_{j_r}+
\frac{d}{dt}\left(\sum\limits_{j_1<\ldots<j_{2n-1}}\prod\limits_{r=1}^{2n-1}\dot{a}_{j_r}\right)\right]\right\}\e^{\sum\limits_ia_i}}
\eq{=\left(\mathfrak{S}_{n}-2n^2\mathfrak{S}_{n-1}\right)\e^{\sum\limits_ia_i}\sum\limits_{j_1<\ldots<j_{2n}}\prod\limits_{r=1}^{2n}\dot{a}_{j_r}+
\frac{d}{dt}\left(\e^{\sum\limits_ia_i}\sum\limits_{j_1<\ldots<j_{2n-1}}\prod\limits_{r=1}^{2n-1}\dot{a}_{j_r}\right),\quad
\mathfrak{S}_n=\prod\limits_{p=0}^{n-1}C^2_{2n-2p}\label{Rn.reduced}}
The last term in the rhs of Eq.~(\ref{Rn.reduced}) is a total time derivative of some function; this term does not contribute to an equation of motion and
can be omitted. Let us denote
\eq{\mathcal{R}_n^{\rm
mod}=2^n\left(\mathfrak{S}_{n}-2n^2\mathfrak{S}_{n-1}\right)\sum\limits_{j_1<\ldots<j_{2n}}\prod\limits_{r=1}^{2n}\dot{a}_{j_r},\quad\mathcal{L}^{\rm
mod}=\frac{1}{16\pi}\sqrt{|g|}\sum\limits_{n=0}^{d}c_n\mathcal{R}_n^{\rm mod}\label{Rn.mod,lagrangian.mod}}
The superscript 'mod' means 'modified'. Using modified Lagrangian~(\ref{Rn.mod,lagrangian.mod}) instead of initial one and varying the
action~(\ref{action}) we obtain dynamical equations
\eq{\sum\limits_{n=1}^{d}\zeta_n\left\{\sum\limits_{k\ne m}\bigl(\ddot{a}_{k}+\dot{a}_k^2\bigr)\sum\limits_{\{j_1<\ldots<j_{2n-2}\}\ne
k,m}\prod\limits_{r=1}^{2n-2}\dot{a}_{j_r}+(2n-1)\sum\limits_{\{j_1<\ldots<j_{2n}\}\ne m}\prod\limits_{r=1}^{2n}\dot{a}_{j_r}\right\}=\kappa^2
T^{m}_{m}\label{Lovelock.equations.final}}
and constraint
\eq{\sum\limits_{n=1}^{d}(2n-1)\zeta_n\sum\limits_{j_1<\ldots<j_{2n}}\prod\limits_{r=1}^{2n}\dot{a}_{j_r}=\kappa^2 T^{0}_{0}\,,\label{constraint}}
where $\zeta_n=c_n2^n\left(\mathfrak{S}_{n}-2n^2\mathfrak{S}_{n-1}\right)$. In the following we consider isotropic perfect fluid with the equation of
state
$p=\omega\rho$ as a matter source, so the energy-momentum tensor takes the form
\eq{T^0_0=-\rho,\quad T^1_1=\ldots=T^D_D=p}
and seek for exponential solutions such that
\eq{ds^2=-dt^2+\sum\limits_{k}\e^{2H_kt}dx_k^2,\quad H_k\equiv\const\label{g.exp}}
Using the notations of the Sec.~\ref{setup} we see that $a_k(t)=H_kt,\;H_k\equiv\const$. Substituting it
in~(\ref{Lovelock.equations.final})--(\ref{constraint}) we see
\eq{\sum\limits_{n=1}^{d}\zeta_n\left\{\sum\limits_{k\ne m}H_k^2\sum\limits_{\{j_1<\ldots<j_{2n-2}\}\ne k,m
}\prod\limits_{r=1}^{2n-2}H_{j_r}+(2n-1)\sum\limits_{\{j_1<\ldots<j_{2n}\}\ne
m}\prod\limits_{r=1}^{2n}H_{j_r}\right\}=\omega\varkappa\label{eqs.of.motion.via.the.Hubble.param}}
\eq{\sum\limits_{n=1}^{d}(2n-1)\zeta_n\sum\limits_{j_1<\ldots<j_{2n}}\prod\limits_{r=1}^{2n}H_{j_r}=-\varkappa,
\quad\varkappa=\kappa^2\rho\label{constr.via.the.Hubble.param}}
It could be useful to rewrite Eqs.~(\ref{eqs.of.motion.via.the.Hubble.param})--(\ref{constr.via.the.Hubble.param}) in terms of elementary symmetric
polynomials (see e.g.~\cite{mcdonald}):
\eq{e_0\equiv 1,\quad e_n=\sum\limits_{j_1<\ldots<j_{n}}\prod\limits_{r=1}^{n}H_{j_r},\quad e_{-n}\equiv 0,\quad n\in\mathbb{N}\label{symm.func}}
In what follows we will also use helpful notation for $e_n$ with parameters $H_{k_1},\ldots,H_{k_l}$ excluded:
\eq{e_n^{k_1,\ldots,k_l}=\sum\limits_{\{j_1<\ldots<j_{n}\}\ne\{k_1,\ldots,k_l\}}\prod\limits_{r=1}^{n}H_{j_r},\quad
n\in\mathbb{N}\label{symm.func.with.excep}}
It is easy to check that
\eq{e_1e_{2n-1}=\sum\limits_i H_i\sum\limits_{j_1<\ldots<j_{2n}}\prod\limits_{r=1}^{2n}H_{j_r}=\sum\limits_{i} H_i^2
\sum\limits_{\{j_1<\ldots<j_{2n-2}\}\ne i} \prod\limits_{r=1}^{2n-2}H_{j_r} + 2n
\sum\limits_{j_1<\ldots<j_{2n}}\prod\limits_{r=1}^{2n}H_{j_r}\label{e1.e2n-1}}
Then with~(\ref{symm.func})--(\ref{e1.e2n-1}) taken into account Eqs.~(\ref{eqs.of.motion.via.the.Hubble.param})--(\ref{constr.via.the.Hubble.param}) read
as
\eq{0=\E_i\equiv\sum\limits_{n=1}^{d}\zeta_n\(e_1^ie_{2n-1}^i-e_{2n}^i\)-\omega\varkappa,\;\;i=1,\ldots,D\,;\qquad
\sum\limits_{n=1}^{d}(2n-1)\zeta_ne_{2n}=-\varkappa\label{final.EoM}}
Since we assume $H_i\equiv\const$, it follows from Eqs.~(\ref{eqs.of.motion.via.the.Hubble.param})--(\ref{constr.via.the.Hubble.param}) that
$\rho\equiv\const$, so that the continuity equation
\eq{\dot\rho+(\rho+p)\sum_i H_i=0}
reduces to
\eq{(\rho+p)\sum_i H_i=0,}
which allows several different cases: a) $\rho\equiv0$ (vacuum case), b) $\rho+p=0$ ($\Lambda$-term case), c) $\sum_i H_i=0$ (constant volume case which
we
call CVS for brevity) and their combinations: d) vacuum CVS and e) $\Lambda$-term CVS. So, all possible exponential solutions can be divided into two
large
groups: solutions with constant volume and solutions with volume changing in time (non-constant volume solutions); for the latter case we have only two
possibilities: vacuum case and $\Lambda$-term case; on the contrary, the first case does not impose constraints on choice of matter a-priory.

Let $N\in\mathbb{N}$ and let $\bgsb{N}$ be the set of the first $N$ natural numbers:
            \eq{\bgsb{N}\equiv\br{l\in\mathbb{N}\bigl|\;l\leqslant N}}
Initial system of $D$ dynamical equations is equivalent to a system that consist of $D-1$ difference equations and one dynamical equation:
\eq{\forall\,i\in\[D\]\;\Bgp{\E_i=0}\iff\forall\,i\in\[D-1\]\;\Bgp{\E_{i+1}-\E_{i}=0}\;\wedge\;\exists\,m\in\[D\]\;\Bgp{\E_m=0}\label{step.1}}
It is easy to check that
\eq{\E_{i_1}-\E_{i_2}=e_1(H_{i_2}-H_{i_1})\sum\limits_{n=1}^{d}\zeta_n e_{2n-2}^{i_1,i_2}=0\label{difference.eq.1}}
Let us introduce the following notations:
\eq{\c_{i_1,i_2}\equiv H_{i_1}-H_{i_2},\qquad\cc_{j_1,\ldots,j_{q}}\equiv\sum\limits_{n=1}^{d}\zeta_n e_{2n-q}^{j_1,\ldots,j_{q}}\label{c1.c2}}
Then we have
\eq{\E_{i+1}-\E_{i}=e_1\,\c_{i,i+1}\,\cc_{i,i+1}=0\iff\c_{i,i+1}=0\quad\vee\quad\cc_{i,i+1}=0\quad\vee\quad\sum_k H_k=0}
Equation $\sum_k H_k=0$ leads to so called constant volume solution (see~\cite{CST1,CST2} for details) and requires separate consideration, so in what
follows we deal with equations $\c_{i,i+1}=0$ and $\cc_{i,i+1}=0$ only.

One more thing needs to be explained before continuing with the general scheme, namely, the coupling constants $\zeta_n$. In previous papers dedicated to
study of the exact solutions in EGB
and Lovelock gravity~\cite{CST1, CST2, prd09, grg10, PT, KPT} we used different couplings constant and for a reader's convenience in comparing solutions we
write down their relation (remember,
$\zeta_n=c_n2^n\left(\mathfrak{S}_{n}-2n^2\mathfrak{S}_{n-1}\right)$ with $\mathfrak{S}_{n}$ given in (\ref{Rn.reduced})):
\eq{\zeta_1 = -2, \zeta_2= -8\alpha, \zeta_3=-144\beta.  \label{add.relation} }
In the absence of $\zeta_2$ and $\zeta_3$ natural normalization for dimensional units would be
\eq{- \zeta_1 = 8\pi G ~\Rightarrow~ 4\pi G \equiv 1,  \label{add.norm} }
\noindent so that when we put some dimensional numerical values, we do it with (\ref{add.norm}) in mind.

\section{The general scheme.}
In this section we describe the general scheme that allows to obtain a wide class of solutions with isotropic subspaces in the framework of general
Lovelock model in $(D+1)$ dimensions.

Let $m\in\z,\,N\in\mathbb{N};\,m\leqslant N$. Let us introduce the following definitions:
\begin{enumerate}
  \item Let $\i^m_{[N]}$ be a collection of all $m$-element subsets of the set $\bgsb{N}$:
            \eq{\i^m_{[N]}=\br{Z\subseteq\bgsb{N}\,\Bigl|\;|Z|=m}\label{k-tuple}}
  Notation $|Z|$ in~(\ref{k-tuple}) stands for cardinality of a set $Z$.
  \item Let $r,k_r,l_{r-1}\in\mathbb{N},\,l_0\equiv D,\,\x_r\in\i^{k_r}_{l_{r-1}-1}$; let us introduce the following notations:
  \eq{\begin{array}{c}
      \j_0\equiv\[D\],\quad\j_k=\j_{k-1}\backslash\widehat{\j}_{k-1},\quad
      \widehat{\j}_{k-1}=\widehat{\j}_{k-1}^1\cup\widehat{\j}_{k-1}^2,\;\;\mbox{where}\\
      \widehat{\j}_{k-1}^1=\br{j_{s}\,\Bigl|\;j_{s}\in\j_{k-1}\;\wedge\;s\in\x_{k}\;\wedge\;s\leqslant\left\lfloor\frac{l_{k-1}-1}{2}\right\rfloor}\\
      \widehat{\j}_{k-1}^2=\br{j_{s+k}\,\Bigl|\;j_{s+k}\in\j_{k-1}\;\wedge\;\;s\in\x_{k}\;\wedge\;s>\left\lfloor\frac{l_{k-1}-1}{2}\right\rfloor}
      \end{array}}
  In what follows we use special notation $y^k_s$ for elements of $\j_k$, i.e. if we write $y^k_s$ somewhere it means there exists a set $\j_k$ such
  that
  $y_s^k\in\j_k$.
\end{enumerate}

{\bf 1st step.} For a given solution some of equations $\E_{i+1}-\E_{i}=0$ are satisfied due to $\c_{i,i+1}=0$, others are satisfied due to
$\cc_{i,i+1}=0$, thus the initial system becomes equivalent to a system that consist of several (say, $k_1$) equations $\c_{i,i+1}=0$, several (say,
$l_1$)
equations $\cc_{i,i+1}=0$ and one dynamical equation:
\eq{
\forall\,i\in\[D\]\;\Bgp{\E_i=0} \iff
\left\{\begin{array}{l}
\exists\,k_1\in\z\;\exists\,\x_1\in\i^{k_1}_{\[D-1\]}\;\forall\,i\in\x_1\;\Bgp{\c_{i,i+1}=0} \\
\exists\,l_1\in\z\;\exists\,\y_1\in\i^{l_1}_{\[D-1\]}\;\forall\,j\in\y_1\;\Bgp{\cc_{j,j+1}=0} \\
\x_1\cap\y_1=\varnothing\;\wedge\;\x_1\cup\y_1=\[D-1\] \\
\E_1=0 \\
k_1+l_1=D-1
\end{array}\right.\label{after.1st.step}}
Note that if $k_1=0$ ($l_1=0$) then $\x_1=\varnothing$ ($\y_1=\varnothing$).

{\bf 2nd step.} The idea is to consider equations $\cc_{j,j+1}=0,\;j\in\y_1$ as a new basic equations, find the respective difference equations and obtain
result analogous to~(\ref{after.1st.step}). The difference $\cc_{j_1,j_2}-\cc_{j_3,j_4}$ is factorized iff one of the elements of the pair $(j_3,j_4)$
equals to one of the elements of the pair $(j_1,j_2)$; for example, let us assume that $j_4=j_2$, then
\eq{\cc_{j_1,j_2}-\cc_{j_2,j_3}=\c_{j_1,j_3}\cc_{j_1,j_2,j_3}\label{diff.2}}
According to~(\ref{c1.c2}) $\c_{i',i'+1}=0\iff H_{i'}=H_{i'+1}$, therefore we can identify indices $i',i'+1$; it is easy to check that one can always find
such $i'\in\x_1,\;j'\in\y_1$ that $i'=j'+1$ (or $i'+1=j'$); using these facts we replace the set $\y_1$ in~(\ref{after.1st.step}) by $\j_1$. By analogy
with~(\ref{step.1}) we have
\eq{\begin{array}{c}
\forall\,k\in\[l_1\]\;\(\cc_{y^1_k,y^1_{k+1}}=0\) \\
\Updownarrow \\
\forall\,k\in\[l_1-1\]\;\(\cc_{y^1_k,y^1_{k+1}}-\cc_{y^1_{k+1},y^1_{k+2}}=0\)\;\wedge\;\exists\,k^1_0\in\[l_1\]\;\(\cc_{y^1_{k^1_0},y^1_{k^1_0+1}}=0\)
\end{array}}
It follows from~(\ref{diff.2}) that
\eq{\cc_{y^1_k,y^1_{k+1}}-\cc_{y^1_{k+1},y^1_{k+2}}=0\iff\c_{y^1_k,y^1_{k+2}}=0\;\vee\;\cc_{y^1_k,y^1_{k+1},y^1_{k+2}}=0\label{diff}}
Some of equations $\cc_{y^1_k,y^1_{k+1}}-\cc_{y^1_{k+1},y^1_{k+2}}=0$ are satisfied due to $\c_{y^1_k,y^1_{k+2}}=0$, others are satisfied due to
$\cc_{y^1_k,y^1_{k+1},y^1_{k+2}}=0$, so that
\eq{\forall\,k\in\[l_1\]\;\Bgp{\cc_{y^1_k,y^1_{k+1}}=0} \iff
\left\{\begin{array}{l}
\exists\,k_2\in\z\;\exists\,\x_2\in\i^{k_2}_{\[l_1-1\]}\;\forall\,p\in\x_2\;\(\c_{y^1_p,y^1_{p+2}}=0\) \\
\exists\,l_2\in\z\;\exists\,\y_2\in\i^{l_2}_{\[l_1-1\]}\;\forall\,q\in\y_2\;\(\cc_{y^1_q,y^1_{q+1},y^1_{q+2}}=0\) \\
\exists\,k^1_0\in\[l_1\]\;\(\cc_{y^1_{k^1_0},y^1_{k^1_0+1}}=0\) \\
\x_2\cap\y_2=\varnothing\;\wedge\;\x_2\cup\y_2=\[l_1-1\] \\
k_2+l_2=l_1-1
\end{array}\right.}
Taking into account 1st and 2nd steps we see that
\eq{\left\{
\begin{array}{c}
  \E_1=0 \\
  \ldots \\
  \E_D=0
\end{array}
\right.
\iff
\left\{
\begin{array}{l}
\c_{i,i+1}=0\;\;\mbox{for all}\;i\in\x_1\in\i^{k_1}_{\[D-1\]}\\
\c_{y^1_p,y^1_{p+2}}=0,\;\;\mbox{for all}\;p\in\x_2\in\i^{k_2}_{\[l_1-1\]}\\
\cc_{y^1_q,y^1_{q+1},y^1_{q+2}}=0,\;\;\mbox{for all}\;q\in\y_2\in\i^{l_2}_{\[l_1-1\]}\\
\cc_{y^1_{k^1_0},y^1_{k^1_0+1}}=0\;\;\mbox{for some}\;k^1_0\in\[l_1\]\\
\E_m=0\;\;\mbox{for some}\;m\in\[D\]\\
k_1+k_2+l_2=D-2
\end{array}
\right.}

{\bf r-th step.} Continuing this procedure at r-th step we have
\eq{\left\{
\begin{array}{c}
  \E_1=0 \\
  \ldots \\
  \E_D=0
\end{array}
\right.
\iff
\left\{
\begin{array}{l}
\c_{p_1,p_1+1}=0\;\;\mbox{for all}\;p_1\in\x_1\in\i^{k_1}_{\[D-1\]}\\
\c_{y^1_{p_2},y^1_{p_2+2}}=0\;\;\mbox{for all}\;p_2\in\x_2\in\i^{k_2}_{\[l_1-1\]}\\
\ldots\ldots \\
\c_{y^{r-1}_{p_r},y^{r-1}_{p_r+r}}=0\;\;\mbox{for all}\;p_r\in\x_r\in\i^{k_r}_{\[l_{r-1}-1\]}\\
\cc_{y^{r-1}_{q_r},\ldots,y^{r-1}_{q_r+r}}=0,\;\;\mbox{for all}\;q_r\in\y_r\in\i^{l_r}_{\[l_{r-1}-1\]},\;l_r\geqslant1\\
\cc_{y^1_{k^1_0},y^1_{k^1_0+1}}=0\;\;\mbox{for some}\;k^1_0\in\[l_1\]\\
\ldots\ldots \\
\cc_{y^{r-1}_{k^{r-1}_0},\ldots,y^{r-1}_{k^{r-1}_0+r-1}}=0\;\;\mbox{for some}\;k^{r-1}_0\in\[l_{r-1}\]\\
\E_m=0\;\;\mbox{for some}\;m\in\[D\]\\
k_1+\ldots+k_r+l_r=D-r
\end{array}
\right.\label{r-th step}}
The process terminates when $l_r\leqslant1$ for some $r=r^*\in\mathbb{N}$; the initial system becomes equivalent to the system consisting of
$k_1+\ldots+k_{r^*}=D-r^*$ equalities of Hubble parameters (equations $\c_{y^i_{p_i},y^i_{p_i+2}}=0$) and $r^*$ additional conditions. The maximal value
of
$r^*$ is $(2d-1)$. Indeed, when $r^*=r^*_{max}\equiv(2d-1)$ degree of the last symmetric polynomial in the sum~(\ref{c1.c2}) becomes zero and
$\cc_{j_1\ldots j_{r^*_{max}}}\equiv\zeta_{d}$. Equation $\cc_{j_1\ldots j_{r^*_{max}}}\equiv\zeta_{d}=0$ means vanishing of the highest Lovelock term in
the Lagrangian and must be omitted, in other words one must set $l_{r^*_{max}}\equiv\nolinebreak0,\;k_{r^*_{max}}\equiv l_{r^*_{max}-1}-1$. Thus, $r^*$
can
vary from 1 to $r^*_{max}$ and, respectively, $(k_1+\ldots+k_{r^*})$ varies from $(D-1)$ to 1 for even $D$ and to 2 for odd $D$.

Each collection of $(k_1+\ldots+k_{r^*})$ equalities of Hubble parameters gives a number of splittings of space into isotropic subspaces. Splittings
corresponding to various values of $(k_1+\ldots+k_{r^*})$ are represented in the Table~\ref{splittings}. We see that even-dimensional spaces has one more
splitting as compared with odd-dimensional one. Numbers in the column "Splitting" means numbers of equal Hubble parameters; braces stand for pairing of
Hubble parameters that give rise to the next splittings; subscripts after round brackets are used to indicate the number of units in these brackets. For
example, record $\lefteqn{\underbrace{\phantom{3+2}}}3+\overbrace{2+(\lefteqn{\underbrace{\phantom{1+1}}}1}+1+\ldots+1)_{D-5}$ means that
$\bgbr{H_1,\ldots,H_D}=\bgbr{H,H,H,h,h,H_6,\ldots,H_D},\;H_6\ne\ldots\ne H_D$ and at the next step one can obtain the following splittings:
\eq{5+(1+\ldots+1)_{D-5},\quad 3+3+(1+\ldots+1)_{D-6},\quad 3+2+2+(1+\ldots+1)_{D-7}}

\begin{table}[t]
\centering
\begin{tabular}{cc|c|cc}
  \cline{2-4}
  & $k_1+\ldots+k_{r^*}$ & Splitting & Number of additional conditions \\
  \cline{2-4}
  \ldelim\{{12}{14.5mm}[even D ]& 1 & $\lefteqn{\underbrace{\phantom{2+(1}}}2+(\overbrace{1+1}+\ldots+1)_{D-2}$ & D-1 &
  \multirow{4}{*}{\rdelim\}{11}{2mm}[
  odd D]} \\
  \cline{2-4}
  & 2 &  $\begin{array}{c}
               \lefteqn{\underbrace{\phantom{3+(1}}}3+(\overbrace{1+1}+\ldots+1)_{D-3} \\
               \lefteqn{\underbrace{\phantom{2+2}}}2+\overbrace{2+(\lefteqn{\underbrace{\phantom{1+1}}}1}+1+\ldots+1)_{D-4}
             \end{array}$ & D-2 &  \\
  \cline{2-4}
  & 3 & $\begin{array}{c}
               \lefteqn{\underbrace{\phantom{4+(1}}}4+(\overbrace{1+1}+\ldots+1)_{D-4} \\
               \lefteqn{\underbrace{\phantom{3+2}}}3+\overbrace{2+(\lefteqn{\underbrace{\phantom{1+1}}}1}+1+\ldots+1)_{D-5}\\
               \underbrace{2+2}+\lefteqn{\overbrace{\phantom{2+(1}}}2+(\underbrace{1+1}+\ldots+1)_{D-6}
             \end{array}$ & D-3 & \\
  \cline{2-4}
  & \ldots\ldots & \ldots\ldots & \ldots\ldots & \\[.8ex]
  & \ldots\ldots & \ldots\ldots & \ldots\ldots & \\
  \cline{2-4}
  & D-1 & isotropic & 1 & \\
  \cline{2-4}
\end{tabular}\\
\caption{Several splittings of D-dimensional space.}
\label{splittings}
\end{table}

\section{Application of the general scheme.\label{application}}
In this section we consider application of the of the scheme described above to the theory with three Lovelock terms in $(7+1)$ dimensions. Let
\eq{\psi_{12}=\frac{\zeta_1}{\zeta_2},\quad\psi_{32}=\frac{\zeta_3}{\zeta_2}}
\begin{table}[t]
\centering
\begin{tabular}{|c|c|c|}
  \hline
  $k_1+\ldots+k_{r^*}$ & Splitting & Number of additional conditions \\
  \hline
  2 &  $\begin{array}{c}
               3+1+1+1+1 \\
               2+2+1+1+1
             \end{array}$ & 5 \\
  \hline
  3 & $\begin{array}{c}
               (4+1+1+1) \\
               (3+2+1+1) \\
               (2+2+2+1)
             \end{array}$ & 4 \\
  \hline
  4 & $\begin{array}{c}
               (5+1+1),\;\;(3+2+2) \\
               (3+3+1),\;\;(4+2+1)
             \end{array}$ & 3 \\
  \hline
  5 & $\begin{array}{c}
               (6+1),\;\;(5+2),\;\;(4+3) \\
             \end{array}$ & 2 \\
  \hline
  6 & isotropic & 1 \\
  \hline
\end{tabular}
\caption{All possible splittings of 7D space.}
\label{splittings-7+1}
\end{table}
\textbf{I.} $\mathbf{\underline{r^*=r^*_{max}=5}};\;l_{5}\equiv0,\,k_1+\ldots+k_5=2$. It is easy to check that
\eq{\begin{array}{c}
         k_1+\ldots+k_5=2 \\
         \Updownarrow \\
         \textbf{1.}\; (k_5=2,l_5=0);\;(k_4=0,l_4=3);\;(k_3=0,l_3=4);\;(k_2=0,l_2=5);\;(k_1=0,l_1=6),\;\mbox{or} \\
         \textbf{2.}\; (k_5=1,l_5=0);\;(k_4=0,l_4=2);\;(k_3=0,l_3=3);\;(k_2=0,l_2=4);\;(k_1=1,l_1=5),\;\mbox{or} \\
         \textbf{3.}\; (k_5=1,l_5=0);\;(k_4=0,l_4=2);\;(k_3=0,l_3=3);\;(k_2=1,l_2=4);\;(k_1=0,l_1=6),\;\mbox{or} \\
         \textbf{4.}\; (k_5=1,l_5=0);\;(k_4=0,l_4=2);\;(k_3=1,l_3=3);\;(k_2=0,l_2=5);\;(k_1=0,l_1=6),\;\mbox{or} \\
         \textbf{5.}\; (k_5=1,l_5=0);\;(k_4=1,l_4=2);\;(k_3=0,l_3=4);\;(k_2=0,l_2=5);\;(k_1=0,l_1=6)\phantom{,}\;\mbox{\phantom{or}} \\
    \end{array}\label{k_1+k_5=2}}
We consider subcases $\textbf{(1)}$ and $\textbf{(2)}$ in more details; other subcases can be considered analogously.

\textbf{(1)} It is easy to check that
\eq{\begin{array}{c}
      \x_1=\varnothing,\;\y_1=\[6\],\;\;\x_2=\varnothing,\;\y_2=\[5\],\;\;\x_3=\varnothing,\;\y_3=\[4\],\;\;\x_4=\varnothing,\;\y_4=\[3\],\;\;
      \x_5=\br{1,2},\;\y_5=\varnothing \\
      \j_1=\j_2=\j_3=\j_4=\[7\]
    \end{array}}
Without loss of generality we can choose $k^1_0=\ldots=k^{4}_0=m=1$, then
\eq{\left\{
\begin{array}{c}
  \E_1=0 \\
  \ldots \\
  \E_7=0
\end{array}
\right.
\iff
\left\{
\begin{array}{l}
\c_{16}=\c_{27}=0 \\
\cc_{12345}=0 \\
\cc_{1234}=0 \\
\cc_{123}=0 \\
\cc_{12}=0 \\
\E_1=0
\end{array}
\right.
\iff
\left\{
\begin{array}{l}
H_1=H_6\equiv x,\;H_2=H_7\equiv y\\
H_3\equiv z,\;H_4\equiv u,\;H_5\equiv v \\
x=-y,\quad\zeta_2-\zeta_3x^2=0 \\
\zeta_2(u+v)-\zeta_3(u+v)x^2=0 \\
\zeta_1+\zeta_2(zu+zv+uv-x^2)-\zeta_3(zu+zv+uv)x^2=0 \\
\sum\limits_{n=1}^{3}\zeta_n\(e_1^1e_{2n-1}^1-e_{2n}^1\)=\omega\varkappa\quad(*)
\end{array}
\right.\label{2+2+1+1+1}}
We see that~(\ref{2+2+1+1+1}) gives us $(2+2+1+1+1)$ splitting and necessary conditions for a set $(x,x,y,y,z,u,v)$ to be a solution of system under
consideration:
\eq{x^2=\psi^{-1}_{32},\,\psi_{32}>0,\,\psi_{12}\psi_{32}=1,\,y=-x;\,z,u,v\in\mathbb{R}}
It is easy to check that the same relations are obtained for any other choice of $k^1_0,\ldots,k^{4}_0$; to get final solution one should substitute these
relations into dynamical equation $(*)$ and constraint~(\ref{constr.via.the.Hubble.param}).
\textbf{(2)} We have
\eq{\begin{array}{c}
      \x_1\in\i_6^1,\;\mbox{let}\,\,\x_1=\br{1},\,\mbox{then}\,\,\y_1=\br{2,\ldots,6},\;\j_1=\br{2,\ldots,7} \\
      \x_2=\varnothing,\;\y_2=\[4\],\;\;\x_3=\varnothing,\;\y_3=\[3\],\;\;\x_4=\varnothing,\;\y_4=\[2\],\;\;\x_5=\br{1},\;\y_5=\varnothing \\
      \j_4=\j_3=\j_2=\j_1=\br{2,\ldots,7}
    \end{array}}
Without loss of generality we can choose $k^1_0=\ldots=k^{4}_0=m=1$, then
\eq{\left\{
\begin{array}{c}
  \E_1=0 \\
  \ldots \\
  \E_7=0
\end{array}
\right.
\iff
\left\{
\begin{array}{l}
\c_{12}=\c_{27}=0\\
\cc_{23}=\cc_{234}=\cc_{2345}=\cc_{23456}=0\\
\E_1=0\\
\end{array}
\right.
\iff
\left\{
\begin{array}{l}
  H_1=H_2=H_7\equiv x \\
  H_3\equiv y,\;H_4\equiv z,\;H_5\equiv u,\;H_6\equiv z,\; \\
  x=0,\;\zeta_2=0 \\
\end{array}\right.
\label{r=5}}
Equality $\zeta_2=0$ implies that the second Lovelock term is absent which contradicts to formulation of the problem (see Sec.~\ref{setup}). So, in the
case under consideration splitting $(3+1+1+1+1)$ is not implemented. It is not difficult to verify that the same result is obtained for any other choice
of
$\x_1$ and $k^1_0,\ldots,k^{4}_0$. One can easily make sure that the rest subcases $\textbf{(3)-(5)}$ does not give us any new results.

\textbf{II.} $\mathbf{\underline{r^*=4}}$. In this case $k_1+\ldots+k_4+l_4=3$; there are a lot of combinations like~(\ref{k_1+k_5=2}) which give such
sum,
so we consider only three of them; one can easily check that another combinations give the same results.

\textbf{(1)} $(k_1=0,l_1=6),\,(k_2=0,l_2=5),\,(k_3=0,l_3=4),\,(k_4=3,l_4=0)$. It easy to check that
\eq{\begin{array}{c}
      \x_1=\varnothing,\;\y_1=\[6\],\;\;\x_2=\varnothing,\;\y_2=\[5\],\;\;\x_3=\varnothing,\;\y_3=\[4\],\;\;\x_4=\br{1,2,3},\;\y_4=\varnothing \\
      \j_1=\j_2=\j_3=\j_4=\[7\]
    \end{array}}
Without loss of generality we can choose $k^1_0=\ldots=k^{3}_0=m=1$, then
\eq{\left\{
\begin{array}{c}
  \E_1=0 \\
  \ldots \\
  \E_7=0
\end{array}
\right.
\iff
\left\{
\begin{array}{l}
\c_{15}=0 \\
\c_{26}=\c_{37}=0 \\
\cc_{1234}=0 \\
\cc_{123}=0 \\
\cc_{12}=0 \\
\E_1=0\\
\end{array}
\right.
\iff
\left\{
\begin{array}{l}
  H_1=H_5\equiv x,\;H_2=H_6\equiv y,\;H_3=H_7\equiv z,\;H_4\equiv u \\
  \zeta_2+\zeta_3(xy+xz+yz)=0 \\
  \zeta_2(x+y+z+u)+\zeta_3(u[xy+xz+yz]+xyz)=0 \\
  \zeta_1+\zeta_2(u[x+y+2z]+xy+2xz+2yz+z^2)+ \\
  \hspace{.8cm}+\zeta_3(u[2xyz+xz^2+yz^2]+xyz^2)=0 \\
  \sum\limits_{n=1}^{3}\zeta_n\(e_1^1e_{2n-1}^1-e_{2n}^1\)=\omega\varkappa\quad(**)
\end{array}\right.
\label{r=4-1}}
We see that~(\ref{r=4-1}) gives us $(2+2+2+1)$ splitting and necessary conditions for a set $(x,x,y,y,z,z,u)$ to be a solution of system under
consideration:
\eq{x^2=\psi^{-1}_{32},\quad\psi_{32}>0,\;\psi_{12}\psi_{32}=1,\quad y=-x,\quad z,u\in\mathbb{R}\label{2+2+2+1}}
It is not difficult to verify that the same result is obtained for any other choice of $k^1_0,\ldots,k^{3}_0$; to get final solution one should substitute
relations~(\ref{2+2+2+1}) into dynamical equation $(**)$ and constraint~(\ref{constr.via.the.Hubble.param}).

\textbf{(2)} $(k_1=0,l_1=6),\,(k_2=1,l_2=4),\,(k_3=0,l_3=3),\,(k_4=2,l_4=0)$. We have
\eq{\begin{array}{c}
      \x_1=\varnothing,\;\y_1=[6],\;\j_1=[7] \\
      \x_2\in\i^1_5,\;\mbox{let}\;\x_2=\br{1},\;\mbox{then}\;\y_2=\br{2,\ldots,5},\;\j_2=\br{2,\ldots,7} \\
      \x_3=\varnothing,\;\y_3=[3],\;\j_3=\j_2.\quad\x_4\br{1,2},\;\y_4=\varnothing
    \end{array}}
Without loss of generality we can choose $k^1_0=\ldots=k^{3}_0=m=1$, then
\eq{\left\{
\begin{array}{c}
  \E_1=0 \\
  \ldots \\
  \E_7=0
\end{array}
\right.
\iff
\left\{
\begin{array}{l}
\c_{13}=0 \\
\c_{26}=\c_{37}=0 \\
\cc_{2345}=0 \\
\cc_{234}=0 \\
\cc_{23}=0 \\
\E_1=0\\
\end{array}
\right.
\iff
\left\{
\begin{array}{l}
  H_1=H_3=H_7\equiv x,\;H_2=H_6\equiv y,\;H_4\equiv z,\;H_5\equiv u \\
  \zeta_2+\zeta_3(x^2+2xy)=0 \\
  \zeta_2(2x+y+u)+\zeta_3(u[x^2+2xy]+x^2y)=0 \\
  \zeta_1+\zeta_2\([2x+y](z+u)+x^2+2xy+zu\)+ \\
  \hspace{.7cm}+\zeta_3\(zu[x^2+2xy]+(z+u)x^2y\)=0 \\
  \sum\limits_{n=1}^{3}\zeta_n\(e_1^1e_{2n-1}^1-e_{2n}^1\)=\omega\varkappa\quad(***)
\end{array}\right.
\label{r=4-2}}
We see that~(\ref{r=4-2}) gives us $(3+2+1+1)$ splitting and necessary conditions for a set $(x,x,x,y,y,z,u)$ to be a solution of system under
consideration:
\eq{y^2=\psi_{32}^{-1},\;\psi_{32}>0,\;\psi_{12}\psi_{32}=1,\quad x=-y,\quad z,u\in\mathbb{R}\label{3+2+1+1}}
It is not difficult to verify that the same result is obtained for any other choice of $\x_2$ and $k^1_0,\ldots,k^{3}_0$; to get final solution one should
substitute relations~(\ref{3+2+1+1}) into dynamical equation $(***)$ and constraint~(\ref{constr.via.the.Hubble.param}).

\textbf{(3)} $(k_1=2,l_1=4),\,(k_2=0,l_2=3),\,(k_3=0,l_3=2),\,(k_4=1,l_4=0)$. We have
\eq{\begin{array}{c}
      \x_1\in\i^2_6,\;\mbox{let}\;\x_1=\br{1,2},\;\mbox{then}\;\y_1=\br{3,4,5,6},\;\j_1=\br{3,\ldots,7} \\
      \x_2=\varnothing,\;\y_2=[3],\;\j_2=\j_1,\quad\x_3=\varnothing,\;\y_3=[2],\;\j_3=\j_2=\j_1,\quad\x_4\br{1},\;\y_4=\varnothing
    \end{array}}
Without loss of generality we can choose $k^1_0=\ldots=k^{3}_0=m=1$, then
\eq{\left\{
\begin{array}{c}
  \E_1=0 \\
  \ldots \\
  \E_7=0
\end{array}
\right.
\iff
\left\{
\begin{array}{l}
\c_{12}=\c_{23}=\c_{37}=0 \\
\cc_{3456}=0 \\
\cc_{345}=0 \\
\cc_{34}=0 \\
\E_1=0\\
\end{array}
\right.
\iff
\left\{
\begin{array}{l}
  H_1=H_2=H_3=H_7\equiv x,\\
  H_4\equiv y,\;H_5\equiv z,\;H_6\equiv u \\
  \zeta_2+3\zeta_3x^2=0 \\
  \zeta_2(3x+u)+\zeta_3(3ux^2+x^3)=0 \\
  \zeta_1+\zeta_2\(3x^2+3x(z+u)+zu\)+ \\
  \hspace{.4cm}+\zeta_3\(3x^2zu+x^3(z+u)\)=0 \\
  \sum\limits_{n=1}^{3}\zeta_n\(e_1^1e_{2n-1}^1-e_{2n}^1\)=\omega\varkappa
\end{array}\right.
  \iff\zeta_2=0
\label{r=5}}
Equality $\zeta_2=0$ implies that the second Lovelock term is absent which contradicts to formulation of the problem (see Sec.~\ref{setup}). So, in the
case under consideration splitting $(4+1+1+1)$ is not implemented. It is not difficult to verify that the same result is obtained for any other choice of
$\x_1$ and $k^1_0,\ldots,k^{3}_0$.

Cases $r^*=1,2,3$ can be considered analogously. We do not give detailed computations due to their awkwardness, but in the next section we describe all
possible splittings in detail for $(7+1)$ and $(6+1)$ dimensions.

\section{Non-constant volume solutions in $(6+1)$ and $(7+1)$ dimensions.\label{solutions}}
Below we give non-constant volume solutions in an implicit or an explicit (wherever it is possible) form for the models with three Lovelock's term
($L_1+L_2+L_3$) and two Lovelock's term ($L_1+L_2$) in $(7+1)$ and $(6+1)$ dimensions.

We have two free parameters in EGB model: $\psi_{12}$ and $\varkappa$. Let $\Omega_2$ be a two dimensional space of parameters $(\psi_{12},\varkappa)$; let $\Omega_2^{\rm sol}$ be a subspace of space $\Omega_2$ such that solutions exist iff $(\psi_{12},\varkappa)\in\Omega_2^{\rm sol},\;\dim\(\Omega_2^{\rm sol}\)\leqslant2$; we point out explicitly the dimensionality of subspace $\Omega_2^{\rm sol}$ if $\dim\(\Omega_2^{\rm sol}\)<2$ and highlight the corresponding splittings by boxes.

In the model with three Lovelock's term there exist three free parameters: $\psi_{12},\varkappa,\psi_{32}$. In a similar way we denote via $\Omega_3$ three-dimensional space of parameters $(\psi_{12},\varkappa,\psi_{32})$ and introduce $\Omega_3^{\rm sol}$ -- subspace of space $\Omega_3$
such that solutions exist iff $(\psi_{12},\varkappa,\psi_{32})\in\Omega_3^{\rm sol},\;\dim\(\Omega_3^{\rm sol}\)\leqslant3$; as in the case of EGB model we point out explicitly the dimensionality of subspace $\Omega_3^{\rm sol}$ if $\dim\(\Omega_3^{\rm sol}\)<3$ and highlight the corresponding splittings by boxes.

\subsection{$(7+1)$-dimensional spacetime.}
\nd{\bf 1.} $(2-2+1+1+1)=(x,x,-x-x,y,z,u)$
\eq{\begin{array}{c}
      \boxed{L_1+L_2+L_3} \vspace{.2cm}\\
      \mbox{necessary conditions:} \vspace{.2cm}\\
      x^2=\psi^{-1}_{32},\quad\psi_{32}>0,\;\psi_{12}\psi_{32}=1,\quad y,z,u\in\mathbb{R} \vspace{.2cm}\\
      \mbox{substitution into equation of motion and constraint leads to} \vspace{.2cm}\\
      \omega=-1,\quad\varkappa=\psi^{-1}_{32},\quad\dim\(\Omega_3^{\rm sol}\)=1 \vspace{.2cm}\\
      \begin{array}{l}
      \mbox{This case involve two subcases:} \vspace{.2cm}\\
      \bullet\;\;\mbox{if $y=z$ then $(2-2+1+1+1)$ is transformed into} \vspace{.2cm}\\
      \hspace{1cm}(2-2+2+1)=(x,x,-x,-x,z,z,u) \vspace{.2cm}\\
      \bullet\;\;\mbox{if $y=x$ then $(2-2+1+1+1)$ is transformed into} \vspace{.2cm}\\
      \hspace{1cm}(3-2+1+1)=(x,x,x,-x,-x,z,u)
      \end{array}
    \end{array}\vsep
    \begin{array}{c}
      L_1+L_2 \vspace{3.1cm} \\
      \mbox{No solutions} \vspace{3.2cm}\\
    \end{array}
\label{case_7_1}}
\hsep
{\bf 2.} $(3-3+1)=(x,x,x,-x,-x,-x,y)$
\eq{\mbox{Let us denote}\quad A_{\pm}=\frac{1}{\psi_{32}}\[1\pm\sqrt{1-\psi_{12}\psi_{32}}\]}
\eq{\begin{array}{c}
      \boxed{L_1+L_2+L_3} \vspace{.2cm}\\
      \mbox{necessary conditions:} \vspace{.2cm}\\
      x^2=\left[\begin{array}{l}
            A_+,\;\;\psi_{12}\psi_{32}\leqslant1,\;\psi_{32}>0 \vspace{.2cm}\\
            A_-,\;\;\psi_{12}\psi_{32}\leqslant1,\;\psi_{12}>0
          \end{array}\right.,\quad y\in\mathbb{R} \vspace{.2cm}\\
      \mbox{Substitution into equation of motion and constraint} \\
      \mbox{leads to} \vspace{.2cm}\\
      \begin{array}{l}
      \mbox{{\bf a)}}\;\;\omega = -1,\quad \varkappa=\dac{\zeta_3x^6-3\zeta_1x^2}{2},\quad\dim\(\Omega_3^{\rm sol}\)=2 \vspace{.2cm}\\
      \mbox{{\bf b)}}\;\;\varkappa=0,\quad\psi_{12}\psi_{32}=\dac{3}{4},\quad\dim\(\Omega_3^{\rm sol}\)=1 \vspace{.2cm}\\
      \end{array}
    \end{array}\hspace{.1cm}\vrule\hspace{.1cm}
    \begin{array}{c}
      \boxed{L_1+L_2} \vspace{.8cm}\\
      \mbox{necessary conditions:} \vspace{.2cm}\\
      \omega=-1,\quad x^2=\dac{\psi_{12}}{2},\;\psi_{12}>0,\quad y\in\mathbb{R} \vspace{.2cm}\\
      \mbox{Substitution into equation of} \\
      \mbox{motion and constraint leads to} \vspace{.2cm}\\
      \varkappa=-\dac{3\zeta_1^2}{2\zeta_2},\quad\zeta_1,\zeta_2<0,\quad\dim\(\Omega_2^{\rm sol}\)=1  \vspace{.8cm}\\
    \end{array}
\label{case_7_2}}
\hsep
{\bf 3.} $(3+3+1)=(x,x,x,y,y,y,z)$
\eq{\begin{array}{c}
      \boxed{L_1+L_2+L_3} \vspace{.2cm}\\
      \mbox{necessary conditions:} \vspace{.2cm}\\
      (x+y)^2=\dac{1-\psi_{12}\psi_{32}}{\psi_{32}},\quad xy=-\psi_{32}^{-1},\quad z\in\mathbb{R} \vspace{.2cm}\\
      \(\psi_{12}\psi_{32}\leqslant1\;\wedge\;\psi_{32}>0\)\;\vee\;\(\psi_{12}\psi_{32}\geqslant5\;\wedge\;\psi_{32}<0\) \vspace{.2cm}\\
      \mbox{Substitution into equation of motion and constraint leads to} \vspace{.2cm}\\
      \begin{array}{l}
      \mbox{{\bf a)}}\;\;\omega = -1,\;\;\varkappa=-\dac{\zeta_1(3x^4 + 9x^3y + 11x^2y^2 + 9xy^3 + 3y^4)}{x^2 + 3xy + y^2},\;\;\dim\(\Omega_3^{\rm sol}\)=2
      \vspace{.2cm}\\
      \mbox{{\bf b)}}\;\;\varkappa=0,\;\;\mbox{let}\;\xi=\dac{y}{x}\;\;\mbox{then} \vspace{.2cm}\\
      \phantom{\mbox{{\bf b)}}\;\;} \left\{\begin{array}{c}
                                             3\xi^4+9\xi^3+11\xi^2+9\xi+3=0,\;\;
                                             \xi_1\approx-1.662,\;\xi_2\approx-0.602 \\
                                             x^2=-\dac{1}{\xi\psi_{32}},\;\;y=\xi x,\;\;\dim\(\Omega_3^{\rm sol}\)=1
                                           \end{array}\right. \vspace{.4cm}\\
      \end{array}
    \end{array}\vsep
    \begin{array}{c}
      L_1+L_2 \vspace{3.2cm}\\
      \mbox{No solutions} \vspace{3.3cm}\\
    \end{array}
\label{case_7_3}}
\hsep
{\bf 4.} $(3+2+2)=(x,x,x,y,y,z,z)$
\eq{\begin{array}{c}
      \mbox{Let us denote} \vspace{.2cm}\\
      B_1(x)\equiv\psi_{32}^3x^8+(\psi_{12}\psi_{32}^3-5\psi_{32}^2)x^6+(7\psi_{32}-\psi_{12}\psi_{32}^2)x^4+
      (\psi_{12}^2\psi_{32}^2-3\psi_{12}\psi_{32}-2)x^2+\psi_{12} \vspace{.2cm}\\
      B_2(x,y)=-\zeta_1(9x^4 + 14x^3y + 11x^2y^2 + 4xy^3 + y^4) \vspace{.2cm}\\
      B_3(\xi)=1026\xi^5+6012\xi^4+15078\xi^3+22159\xi^2+16524\xi+5196 \vspace{.2cm}\\
      B_4(x,y,z)=3x^5(y+z)+6x^4(y+z)^2+4x^3(y^3+4y^2z+4yz^2+z^3)+ \\
                   2x^2(y^4+5y^3z+8y^2z^2+5yz^3+z^4)+4x(y^4z+2y^3z^2+2y^2z^3+yz^4)+2y^4z^2 + 2y^3z^3 + 2y^2z^4 \vspace{.2cm}\\
      B_5(x,y,z)=x^3y + x^3z + 2x^2y^2 + 4x^2yz + 2x^2z^2
    \end{array}
}
\eq{\begin{array}{c}
L_1+L_2+L_3 \vspace{.2cm}\\
\mbox{necessary conditions:} \vspace{.2cm}\\
z=\dac{\(\psi_{12}\psi_{32}-1\)x\pm\sqrt{B_1(x)}}{(\psi_{32}x^2-1)^2},\; y=-\dac{\psi_{32}x^2z+2x+z}{\psi_{32}x(x+2z)+1} \vspace{.2cm}\\
\mbox{with $x$ such that}\;\;B_1(x)\geqslant0 \vspace{.2cm}\\
\mbox{Substitution into equation of motion and constraint}\\
\mbox{leads to} \vspace{.2cm}\\
\begin{array}{l}
      \mbox{{\bf a)}}\;\;\omega = -1,\;\;\varkappa=\dac{2B_4(x,y,z)}{B_5(x,y,z)} \vspace{.2cm}\\
      \mbox{{\bf b)}}\;\;\varkappa=0,\;\;\mbox{let}\;\xi=\dac{y}{x},\;\chi=\dac{z}{x}\;\;\mbox{then equation}\;\;B_4(x,y,z)=0 \\
      \phantom{\mbox{{\bf b)}}\;\;\varkappa=0,\;\;\mbox{let}\;\xi=\dac{y}{x},\;\chi=\dac{z}{x}\;\;}\mbox{take the following form:} \vspace{.2cm}\\
      \begin{array}{c}
                                      2\, \left( \xi+1 \right) ^{2}{\chi}^{4}+2\, \left( \xi+2 \right)
                                      \left( \xi+1 \right) ^{2}{\chi}^{3}+2\, \left( {\xi}^{2}+2\,\xi+3\right)\left( \xi+1 \right) ^{2}{\chi}^{2} \\
                                      +\left( 4\,{\xi}^{4}+10\,{\xi}^{3}+16\,{\xi}^{2}+12\,\xi+3 \right)
                                      \chi+2\,{\xi}^{4}+4\,{\xi}^{3}+6\,{\xi}^{2}+3\,\xi=0
                                    \end{array} \vspace{.2cm}\\
      \mbox{Special solution:}\vspace{.2cm}\\
      \phantom{\mbox{{\bf b)}}\;\;} \left\{\begin{array}{c}
                                      x^2=\dac{\psi_{12}}{432}B_3(\xi),\;\;y=\xi x,\;\;z=-\dac{2\xi x}{3\xi+2} \\
                                      18\xi^6+120\xi^5+350\xi^4+603\xi^3+606\xi^2+324\xi+72=0, \\
                                      \xi_1\approx-2.556,\;\xi_2\approx-0.902 \\
                                    \end{array}\right.
      \end{array}
\end{array}
\vrule
\begin{array}{c}
      L_1+L_2 \vspace{2.7cm}\\
      \mbox{necessary conditions:} \vspace{.2cm}\\
      z=-x\pm\sqrt{\psi_{12}-2x^2},\\ y=-2x-z \vspace{.2cm}\\
      \mbox{$x$ such that}\;x^2\leqslant\dac{\psi_{12}}{2},\;\psi_{12}>0 \vspace{.2cm}\\
      \mbox{Substitution into equation of}\\
      \mbox{motion and constraint leads to} \vspace{.2cm}\\
      \omega=-1,\;\varkappa=\dac{B_2(x,y)}{3x^2 + 2xy + y^2} \vspace{2.7cm}\\
\end{array}\label{case_7_4}}
\hsep
{\bf 5.} $(4+2+1)=(x,x,x,x,y,y,z)$
\eq{\begin{array}{c}
      \boxed{L_1+L_2+L_3} \vspace{.3cm}\\
      \mbox{necessary conditions:} \vspace{.2cm}\\
      \psi_{32}^2x^6-3\psi_{32}x^4-3\(\psi_{12}\psi_{32}-2\)x^2-\psi_{12}=0 \vspace{.2cm}\\
      y=-\dac{x(\psi_{32}x^2+3)}{3\psi_{32}x^2+1},\quad z\in\mathbb{R} \vspace{.2cm}\\
      \mbox{Substitution into equation of motion and constraint} \\
      \mbox{leads to}\vspace{.2cm}\\
      \omega = -1,\;\;\varkappa = -\dac{\zeta_1(15x^4 + 45x^3y + 48x^2y^2 + 24xy^3 + 8y^4)}{3x^2 + 9xy + 8y^2} \vspace{.2cm}\\
      \dim\(\Omega_3^{\rm sol}\)=2
    \end{array}\hspace{.2cm}\vrule\hspace{.2cm}
    \begin{array}{c}
      \boxed{L_1+L_2} \vspace{.3cm}\\
      \mbox{necessary conditions:} \vspace{.2cm}\\
      x^2=\dac{\psi_{12}}{6},\;\psi_{12}>0\vspace{.2cm}\\
      y=-3x,\quad z\in\mathbb{R} \vspace{.2cm}\\
      \mbox{Substitution into equation of} \\
      \mbox{motion and constraint leads to} \vspace{.2cm}\\
      \omega=-1,\quad\varkappa=-\dac{13\zeta_1^2}{6\zeta_2},\;\;\zeta_1,\zeta_2<0 \vspace{.2cm}\\
      \dim\(\Omega_2^{\rm sol}\)=1
    \end{array}
\label{case_7_5}}
\hsep
{\bf 6.} $(4+3)=(x,x,x,x,y,y,y)$
\eq{\begin{array}{c}
      \mbox{Let us denote} \vspace{.2cm}\\
      F_1(x)\equiv\psi_{32}^2x^6-3\psi_{32}x^4-3(\psi_{12}\psi_{32}-2)x^2-\psi_{12} \vspace{.2cm}\\
      F_2(x,y)\equiv-\zeta_2(6x^5 + 36x^4y + 66x^3y^2 + 51x^2y^3+16xy^4)-\zeta_1(12x^3 + 27x^2y + 22xy^2 + 9y^3) \vspace{.2cm}\\
      F_3(x,y)\equiv-3\zeta_1(5x^4 + 12x^3y + 11x^2y^2 + 6xy^3 + y^4)
    \end{array}}
\eq{\begin{array}{c}
      L_1+L_2+L_3 \vspace{.6cm}\\
      \mbox{necessary conditions:} \vspace{.2cm}\\
      y=\dac{-\psi_{32}x^3-3x\pm\sqrt{F_1(x)}}{3\psi_{32}x^2+1},\;
      \mbox{$x$ such that}\;\;F_1(x)\geqslant0 \vspace{.2cm}\\
      \mbox{Substitution into equation of motion} \\
      \mbox{and constraint leads to} \vspace{.2cm}\\
      \begin{array}{l}
      \mbox{{\bf a)}}\;\;\omega = -1,\quad\varkappa = \dac{F_2(x,y)}{2x+3y} \vspace{.2cm}\\
      \mbox{{\bf b)}}\;\;\varkappa=0,\quad F_2(x,y)=0 \vspace{1.8cm}\\
      \end{array}
    \end{array}\hspace{.1cm}\vrule\hspace{.1cm}
    \begin{array}{c}
      L_1+L_2 \vspace{.2cm}\\
      \mbox{necessary conditions:} \vspace{.2cm}\\
      3x^2 + 6xy + y^2=-\psi_{12} \vspace{.2cm}\\
      \mbox{Substitution into equation of motion} \\
      \mbox{and constraint leads to} \vspace{.2cm}\\
      \begin{array}{l}
      \mbox{{\bf a)}}\;\;\omega = -1,\quad\varkappa = -\psi_{12}^{-1}F_3(x,y) \vspace{.2cm}\\
      \mbox{{\bf b)}}\;\;\varkappa=0,\;\;\mbox{let}\;\xi=\dac{y}{x}\;\;\mbox{then} \vspace{.2cm}\\
      \phantom{\mbox{{\bf b)}}\;\;} \left\{\begin{array}{c}
                                      \xi^4+6\xi^3+11\xi^2+ 12\xi+5=0, \\
                                      \xi_1\approx-3.874,\;\xi_2\approx-0.743 \\
                                      x^2=-\dac{\psi_{12}}{\xi^2 + 6\xi + 3},\;\;y=\xi x
                                    \end{array}\right. \vspace{.2cm}
      \end{array}
    \end{array}
\label{case_7_6}}
\hsep
{\bf 7.} $(5+1+1)=(x,x,x,x,x,y,z)$
\eq{\begin{array}{c}
      \boxed{L_1+L_2+L_3} \vspace{.2cm}\\
      \mbox{necessary conditions:} \vspace{.2cm}\\
      x^2=-\psi_{32}^{-1},\quad\psi_{32}<0,\;\;\psi_{12}\psi_{32}=5,\quad y,z\in\mathbb{R}\vspace{.2cm}\\
      \mbox{Substitution into equation of motion and constraint leads to} \vspace{.2cm}\\
      \omega = -1,\quad \varkappa = -7\psi_{32}^{-1},\quad\dim\(\Omega_3^{\rm sol}\)=1 \vspace{.2cm}\\
    \end{array}\vsep
    \begin{array}{c}
      L_1+L_2 \vspace{1.2cm}\\
      \mbox{No solutions} \vspace{1.3cm}\\
    \end{array}
\label{case_7_7}}
\hsep
{\bf 8.} $(5+2)=(x,x,x,x,x,y,y)$
\eq{\begin{array}{c}
      \mbox{Let us denote} \vspace{.2cm}\\
      G_1(x,y)=-\zeta_2(15x^5 + 60x^4y + 80x^3y^2 + 20x^2y^3)-\zeta_1(10x^3 + 40x^2y + 16xy^2 + 4y^3) \vspace{.2cm}\\
      G_2(x,y)=-\zeta_1(45x^3 + 40x^2y + 16xy^2 + 4y^3)
    \end{array}
}
\eq{\begin{array}{c}
      L_1+L_2+L_3 \vspace{.8cm}\\
      \mbox{necessary conditions:} \vspace{.2cm}\\
      y=-\dac{\psi_{32}x^4+6x^2+\psi_{12}}{4x(1+\psi_{32}x^2)},\quad x\in\mathbb{R} \vspace{.2cm}\\
      \mbox{Substitution into equation of motion and} \\
      \mbox{constraint leads to} \vspace{.2cm}\\
      \begin{array}{l}
      \mbox{{\bf a)}}\;\;\omega = -1,\quad\varkappa = \dac{G_1(x,y)}{x+4y} \vspace{.2cm}\\
      \mbox{{\bf b)}}\;\;\varkappa=0,\quad G_1(x,y)=0 \vspace{1.3cm}\\
      \end{array}
    \end{array}\hspace{.3cm}\vrule\hspace{.3cm}
    \begin{array}{c}
      L_1+L_2 \vspace{.2cm}\\
      \mbox{necessary conditions:} \vspace{.2cm}\\
      y=-\dac{6x^2+\psi_{12}}{4x},\quad x\in\mathbb{R} \vspace{.2cm}\\
      \mbox{Substitution into equation of motion and} \\
      \mbox{constraint leads to} \vspace{.2cm}\\
      \begin{array}{l}
      \mbox{{\bf a)}}\;\;\omega = -1,\quad\varkappa = \dac{G_2(x,y)}{2(3x+2y)} \vspace{.2cm}\\
      \mbox{{\bf b)}}\;\;\varkappa=0,\;\;\mbox{let}\;\xi=\dac{y}{x}\;\;\mbox{then} \vspace{.2cm}\\
      \phantom{\mbox{{\bf b)}}\;\;} \left\{\begin{array}{c}
                                      4\xi^3+16\xi^2+40\xi+45=0,\;\xi\approx-1.870 \\
                                      x^2=-\dac{\psi_{12}}{2(2\xi + 3)},\;\;y=\xi x
                                    \end{array}\right.
      \end{array}
    \end{array}
\label{case_7_8}}
\hsep
{\bf 9.} $(6+1)=(x,x,x,x,x,x,y)$
\eq{\mbox{Let us denote}\quad J_{\pm}=\dac{1}{\psi_{32}}\[-1\pm\sqrt{1-\dac{\psi_{12}\psi_{32}}{5}}\]}
\eq{\begin{array}{c}
      \boxed{L_1+L_2+L_3} \vspace{.3cm}\\
      \mbox{necessary conditions:} \vspace{.2cm}\\
      x^2=\left[\begin{array}{l}
            J_+,\;\;\psi_{12}\psi_{32}\leqslant5,\;\psi_{12}<0 \vspace{.2cm}\\
            J_-,\;\;\psi_{12}\psi_{32}\leqslant5,\;\psi_{32}<0
          \end{array}\right.,\quad y\in\mathbb{R} \vspace{.2cm}\\
      \mbox{Substitution into equation of motion and} \\
      \mbox{constraint leads to} \vspace{.2cm}\\
      \begin{array}{l}
      \mbox{{\bf a)}}\;\;\omega = -1,\quad\varkappa = \dac{35\zeta_3x^6-21\zeta_1x^2}{2},\quad\dim\(\Omega_3^{\rm sol}\)=2 \vspace{.2cm}\\
      \mbox{{\bf b)}}\;\;\varkappa=0,\quad \psi_{12}\psi_{32}=\dac{15}{4},\quad\dim\(\Omega_3^{\rm sol}\)=1 \vspace{.2cm}\\
      \end{array}
    \end{array}\hspace{.1cm}\vrule\hspace{.1cm}
    \begin{array}{c}
      \boxed{L_1+L_2} \vspace{.3cm}\\
      \mbox{necessary conditions:} \vspace{.2cm}\\
      x^2=\dac{\psi_{12}}{10},\;\psi_{12}>0,\quad y\in\mathbb{R} \vspace{.2cm}\\
      \mbox{Substitution into equation of motion and} \\
      \mbox{constraint leads to} \vspace{.2cm}\\
      \omega = -1,\quad\varkappa=-\dac{21\zeta_1^2}{20\zeta_2},\;\;\zeta_1,\zeta_2<0 \vspace{.2cm}\\
      \dim\(\Omega_2^{\rm sol}\)=1
      \vspace{.5cm}\\
    \end{array}
\label{case_7_9}}
\hsep
{\bf 10.} $(7+0)=(x,x,x,x,x,x,x)$
\eq{\begin{array}{c}
      \mbox{Let us denote} \vspace{.2cm}\\
      K_{\pm}=\dac{3}{2\psi_{32}}\[-1\pm\sqrt{1-\dac{4\psi_{12}\psi_{32}}{15}}\],\quad
      L_{\pm}=\dac{\psi_{12}}{10}\[-1\pm\sqrt{1-\dac{20}{21}\,\dac{\varkappa}{\psi_{12}\zeta_1}}\]
    \end{array}}
\eq{\begin{array}{c}
      L_1+L_2+L_3 \vspace{.6cm}\\
      \mbox{Substitution $H_i\equiv x$ into equation of motion} \\
      \mbox{and constraint leads to} \vspace{.2cm}\\
      \begin{array}{l}
      \mbox{{\bf a)}}\;\;\omega = -1,\quad\varkappa = -35\zeta_3x^6-105\zeta_2x^4-21\zeta_1x^2 \vspace{.2cm}\\
      \mbox{{\bf b)}}\;\;\varkappa=0,\quad
      x^2=\left[\begin{array}{l}
            K_+,\;\;\psi_{12}\psi_{32}\leqslant\dac{15}{4},\;\psi_{12}<0 \vspace{.2cm}\\
            K_-,\;\;\psi_{12}\psi_{32}\leqslant\dac{15}{4},\;\psi_{32}<0
          \end{array}\right.
      \end{array}
      \vspace{1.5cm}\\
    \end{array}\vrule\hspace{.1cm}
    \begin{array}{c}
      L_1+L_2 \vspace{.5cm}\\
      \mbox{Substitution $H_i\equiv x$ into equation of motion} \\
      \mbox{and constraint leads to} \vspace{.2cm}\\
      \begin{array}{l}
      \mbox{{\bf a)}}\;\;\omega = -1,\;
      x^2=\left[\begin{array}{l}
            L_+,\;\;\dac{\varkappa}{\psi_{12}\zeta_1}\leqslant\dac{21}{20},\;\zeta_1<0 \vspace{.2cm}\\
            L_-,\;\;\dac{\varkappa}{\psi_{12}\zeta_1}\leqslant\dac{21}{20},\;\psi_{12}<0
          \end{array}\right.  \vspace{.3cm}\\
      \mbox{{\bf b)}}\;\;\varkappa=0,\quad x^2=-\dac{\psi_{12}}{5},\;\;\psi_{12}<0
      \end{array}
      \vspace{1cm}\\
    \end{array}
\label{case_7_10}}
One can see that in (7+1) dimensions solutions for $L_1+L_2+L_3$ are more abundant then solutions for $L_1+L_2$ (EGB) -- probably due to the higher number
of degrees of freedom -- we will discuss this aspect in detail in the appropriate section. For the reader's convenience we summarize all the solutions in
Table \ref{7.plus.1}.

\begin{table}[!h]
\begin{center}
\caption{Summary of (7+1)-dimensional solutions in \Lall and EGB.}
\label{7.plus.1}
  \begin{tabular}{|c|c|c|c|c|}
    \hline
    Splitting & nonvacuum \Lall & vacuum \Lall & nonvacuum EGB & vacuum EGB  \\
    \hline
    $(2-2+1+1+1)$ (\ref{case_7_1}) & yes & \multicolumn{3}{c|}{no} \\
    \hline
    $(3-3+1)$ (\ref{case_7_2}) & \multicolumn{3}{c|}{yes} & no \\
    \hline
    $(3+3+1)$ (\ref{case_7_3}) & \multicolumn{2}{c|}{yes} & \multicolumn{2}{c|}{no} \\
    \hline
    $(3+2+2)$ (\ref{case_7_4}) & \multicolumn{3}{c|}{yes} & no \\
    \hline
    $(4+2+1)$ (\ref{case_7_5}) & yes & no & yes & no \\
    \hline
    $(4+3)$ (\ref{case_7_6}) & \multicolumn{4}{c|}{yes} \\
    \hline
    $(5+1+1)$ (\ref{case_7_7}) & yes & \multicolumn{3}{c|}{no} \\
    \hline
    $(5+2)$ (\ref{case_7_8}) & \multicolumn{4}{c|}{yes} \\
    \hline
    $(6+1)$ (\ref{case_7_9}) & \multicolumn{3}{c|}{yes} & no \\
    \hline
    $(7+0)$ (\ref{case_7_10}) & \multicolumn{4}{c|}{yes} \\
    \hline
  \end{tabular}
\end{center}
\end{table}

\subsection{(6+1)-dimensional spacetime.}
{\bf 1.} $(2-2+1+1)=(x,x,-x,-x,y,z)$
\eq{\begin{array}{c}
      \boxed{L_1+L_2+L_3} \vspace{.2cm}\\
      \mbox{necessary conditions:} \vspace{.2cm}\\
      x^2=\psi_{32}^{-1},\quad y,z\in\mathbb{R},\quad\psi_{12}\psi_{32}=1 \vspace{.2cm}\\
      \mbox{substitution into equation of motion and constraint leads to} \vspace{.2cm}\\
      \omega=-1,\quad\varkappa=\psi^{-1}_{32},\quad\dim\(\Omega_3^{\rm sol}\)=1 \vspace{.2cm}\\
    \end{array}\vsep
    \begin{array}{c}
      L_1+L_2 \vspace{1.4cm} \\
      \mbox{No solutions} \vspace{1.4cm}\\
    \end{array}
\label{add_case_6_1}}
\hsep
{\bf 2.} $(2+2+2)=(x,x,y,y,z,z)$
\eq{\begin{array}{c}
      \mbox{Let us denote} \vspace{.2cm}\\
      M_1(y)\equiv4\psi_{32}y^4+\bgsb{\psi_{12}^2\psi_{32}^2-6\psi_{12}\psi_{32}-3}y^2+4\psi_{12} \\
      M_2(x,y)\equiv x^3y + x^3z + x^2y^2 + 2x^2yz + x^2z^2 + xy^3 + 2xy^2z + 2xyz^2 + xz^3 + y^3z ~+ y^2z^2 + yz^3
    \end{array}}
\eq{\begin{array}{c}
L_1+L_2+L_3 \vspace{.2cm}\\
\mbox{necessary conditions:} \vspace{.2cm}\\
x=\dac{\bgp{1-\psi_{12}\psi_{32}}y\pm\sqrt{M_1(y)}}{2\bgp{\psi_{32}y^2-1}},\;z=-\frac{x+y}{1+\psi_{32}xy} \vspace{.2cm}\\
\mbox{with $y$ such that}\;\;M_1(y)\geqslant0 \vspace{.2cm}\\
\mbox{Substitution into equation of motion and constraint}\\
\mbox{leads to} \vspace{.2cm}\\
\omega = -1,\quad \varkappa=\dac{M_2(x,y)}{xy + xz + yz}
\end{array}
\vsep
\begin{array}{c}
      L_1+L_2\vspace{2.2cm} \\
      \mbox{No solutions} \vspace{2.2cm}\\
\end{array}\label{add_case_6_2}}
\hsep
{\bf 3.} $(3+2+1)=(x,x,x,y,y,z)$
\eq{\begin{array}{c}
      \mbox{Let us denote} \vspace{.2cm}\\
      N^{\pm}_1\equiv\dac{1}{2\psi_{32}}\[3-\psi_{12}\psi_{32}\pm\sqrt{(3-\psi_{12}\psi_{32})^2-4\psi_{12}\psi_{32}}\],\quad
      N_2(x,y)\equiv-\zeta_1(3x^3+6x^2y+4xy^2+2y^3)
    \end{array}}
\eq{\begin{array}{c}
L_1+L_2+L_3 \vspace{.2cm}\\
\mbox{necessary conditions:} \vspace{.2cm}\\
z\in\mathbb{R},\quad y=-\dac{2x}{1+\psi_{32}x^2} \vspace{.2cm}\\
x^2=\left[\begin{array}{l}
            N^+_1,\;\;\mbox{for}\;
            \left\{\begin{array}{c}
                      \psi_{32}>0 \\
                      \psi_{12}\psi_{32}\leqslant1
                    \end{array}\right.\;\mbox{or}\;
            \left\{\begin{array}{c}
                      \psi_{32}<0 \\
                      \psi_{12}\psi_{32}\geqslant9
                    \end{array}\right.\vspace{.2cm}\\
            N^-_1,\;\;\mbox{for}\;
            \left\{\begin{array}{c}
                      \psi_{32}>0,\;\psi_{12}>0 \\
                      \psi_{12}\psi_{32}\leqslant1
                    \end{array}\right.\;\mbox{or}\;
            \left\{\begin{array}{c}
                      \psi_{32}<0 \\
                      \psi_{12}\psi_{32}\geqslant9\;\;\mbox{or}\;\;\psi_{12}>0
                    \end{array}\right.
          \end{array}\right. \vspace{.2cm}\\
\mbox{Substitution into equation of motion and constraint}\\
\mbox{leads to} \vspace{.2cm}\\
\begin{array}{l}
      \mbox{{\bf a)}}\;\;\omega=-1,\quad\varkappa=\dac{N_2(x,y)}{x+2y} \vspace{.2cm}\\
      \mbox{{\bf b)}}\;\;\varkappa=0,\;\;\mbox{let}\;\xi=\dac{y}{x}\;\;\mbox{then} \vspace{.2cm}\\
      \phantom{\mbox{{\bf b)}}\;\;} \left\{\begin{array}{c}
                                      2\xi^3+4\xi^2+6\xi+3=0,\;\xi\approx-0.722 \\
                                      x^2=-\dac{\xi+2}{\xi\psi_{32}},\;\;y=\xi x
                                    \end{array}\right.
      \end{array}
\end{array}
\vrule\hspace{.1cm}
\begin{array}{c}
      \boxed{L_1+L_2} \vspace{1.2cm}\\
      \mbox{necessary conditions:} \vspace{.2cm}\\
      x^2=\dac{\psi_{12}}{3},\;\psi_{12}>0,\\
      y=-2x,\quad z\in\mathbb{R}  \vspace{.2cm}\\
      \mbox{Substitution into equation} \\
      \mbox{of motion and constraint} \\
      \mbox{leads to} \vspace{.2cm}\\
      \omega = -1,\quad\varkappa=-\dac{2\zeta_1^2}{\zeta_2},\vspace{.2cm}\\
      \zeta_1,\zeta_2<0,\;\;\dim\(\Omega_2^{\rm sol}\)=1 \\
      \vspace{1.5cm}
\end{array}\label{add_case_6_3}}
\hsep
{\bf 4.} $(4+1+1)=(x,x,x,x,y,z)$
\eq{\begin{array}{c}
\boxed{L_1+L_2+L_3} \vspace{.2cm}\\
\mbox{necessary conditions:} \vspace{.2cm}\\
x^2=-3\psi_{32}^{-1},\;\;\psi_{32}<0,\;\;\psi_{12}\psi_{32}=9,\quad y,z\in\mathbb{R} \vspace{.2cm}\\
\mbox{Substitution into equation of motion and constraint}\\
\mbox{leads to} \vspace{.2cm}\\
\omega = -1,\quad \varkappa=-15\psi_{32}^{-1},\quad\dim\(\Omega_3^{\rm sol}\)=1 \vspace{.2cm}\\
\end{array}
\vsep
\begin{array}{c}
      L_1+L_2\vspace{1.6cm} \\
      \mbox{No solutions} \vspace{1.7cm}\\
\end{array}\label{add_case_6_4}}
\hsep
{\bf 5.} $(3+3)=(x,x,x,y,y,y)$
\eq{\begin{array}{c}
      \mbox{Let us denote} \vspace{.2cm}\\
      P_1(x)=-\psi_{32}x^4+(3-\psi_{12}\psi_{32})x^2-\psi_{12} \vspace{.2cm}\\
      P_2(x,y)=-\zeta_2xy(4x^2 + 7xy + 4y^2)-\zeta_1(3x^2 + 4xy + 3y^2) \vspace{.2cm}\\
      P_3(x,y)=-3\zeta_1(x^4 + 4x^3y + 5x^2y^2 + 4xy^3 + y^4)
    \end{array}
}
\eq{\begin{array}{c}
      L_1+L_2+L_3 \vspace{1cm}\\
      \mbox{necessary conditions:} \vspace{.2cm}\\
      y=\dac{-2x\pm\sqrt{P_1(x)}}{\psi_{32}x^2+1},\;
      \mbox{$x$ such that}\;\;P_1(x)\geqslant0 \vspace{.2cm}\\
      \mbox{Substitution into equation of motion} \\
      \mbox{and constraint leads to} \vspace{.2cm}\\
      \begin{array}{l}
      \mbox{{\bf a)}}\;\;\omega = -1,\quad\varkappa = P_2(x,y) \vspace{.2cm}\\
      \mbox{{\bf b)}}\;\;\varkappa=0,\quad P_2(x,y)=0 \vspace{2.2cm}\\
      \end{array}
    \end{array}\hspace{.1cm}\vrule\hspace{.1cm}
    \begin{array}{c}
      L_1+L_2 \vspace{.2cm}\\
      \mbox{necessary conditions:} \vspace{.2cm}\\
      y=-2x\pm\sqrt{3x^2-\psi_{12}}\\
      \mbox{with $x$ such that}\;\;x^2\geqslant\dac{\psi_{12}}{3} \vspace{.2cm}\\
      \mbox{Substitution into equation of motion} \\
      \mbox{and constraint leads to} \vspace{.2cm}\\
      \begin{array}{l}
      \mbox{{\bf a)}}\;\;\omega = -1,\quad\varkappa = \dac{P_3(x,y)}{x^2 + 4xy + y^2} \vspace{.2cm}\\
      \mbox{{\bf b)}}\;\;\varkappa=0,\;\;\mbox{let}\;\xi=\dac{y}{x}\;\;\mbox{then} \vspace{.2cm}\\
      \phantom{\mbox{{\bf b)}}\;\;} \left\{\begin{array}{c}
                                      \xi^4+4\xi^3+5\xi^2+4\xi+1=0,\;\xi=-\dac{3}{2}\pm\dac{\sqrt{5}}{2} \\
                                      x^2=-\dac{\psi_{12}}{\xi^2+4\xi+1},\;\;y=\xi x
                                    \end{array}\right.
      \end{array}
    \end{array}
\label{add_case_6_5}}
\hsep
{\bf 6.} $(4+2)=(x,x,x,x,y,y)$
\eq{\begin{array}{c}
      \mbox{Let us denote} \vspace{.2cm}\\
      Q_1(x,y)=-12\zeta_2(x^4 + 3x^3y + x^2y^2)-\zeta_1(6x^2 + 3xy + y^2) \vspace{.2cm}\\
      Q_2(x,y)=-\zeta_1(5x^3 + 6x^2y + 3xy^2 + y^3) \vspace{.2cm}\\
    \end{array}
}
\eq{\begin{array}{c}
      L_1+L_2+L_3 \vspace{.8cm}\\
      \mbox{necessary conditions:} \vspace{.2cm}\\
      y=-\dac{3x^2+\psi_{12}}{\psi_{32}x^3+3x},\quad x\in\mathbb{R} \vspace{.2cm}\\
      \mbox{Substitution into equation of motion and constraint} \\
      \mbox{leads to} \vspace{.2cm}\\
      \begin{array}{l}
      \mbox{{\bf a)}}\;\;\omega = -1,\quad\varkappa = Q_1(x,y) \vspace{.2cm}\\
      \mbox{{\bf b)}}\;\;\varkappa=0,\quad Q_1(x,y)=0 \vspace{.9cm}\\
      \end{array}\vspace{.8cm}\\
    \end{array}\hspace{.1cm}\vrule\hspace{.1cm}
    \begin{array}{c}
      L_1+L_2 \vspace{.2cm}\\
      \mbox{necessary conditions:} \vspace{.2cm}\\
      y=-\dac{3x^2+\psi_{12}}{3x},\quad x\in\mathbb{R}\vspace{.2cm}\\
      \mbox{Substitution into equation of motion and} \\
      \mbox{constraint leads to} \vspace{.2cm}\\
      \begin{array}{l}
      \mbox{{\bf a)}}\;\;\omega = -1,\quad\varkappa = \dac{Q_2(x,y)}{x+y} \vspace{.2cm}\\
      \mbox{{\bf b)}}\;\;\varkappa=0,\;\;\mbox{let}\;\xi=\dac{y}{x}\;\;\mbox{then} \vspace{.2cm}\\
      \phantom{\mbox{{\bf b)}}\;\;} \left\{\begin{array}{c}
                                      \xi^3+3\xi^2+6\xi+5=0,\;\xi=-1.322 \\
                                      x^2=-\dac{\psi_{12}}{3(\xi+1)},\;\;y=\xi x
                                    \end{array}\right.
      \end{array}
    \end{array}
\label{add_case_6_6}}
\hsep
{\bf 7.} $(5+1)=(x,x,x,x,x,y)$
\eq{\mbox{Let us denote}\quad U_{\pm}=\dac{1}{\psi_{32}}\[-3\pm\sqrt{9-\psi_{12}\psi_{32}}\]}
\eq{\begin{array}{c}
\boxed{L_1+L_2+L_3} \vspace{.2cm}\\
\mbox{necessary conditions:} \vspace{.2cm}\\
y\in\mathbb{R},\quad
x^2=\left[\begin{array}{l}
            U_+,\;\;\mbox{for}\;
            \left\{\begin{array}{c}
                      \psi_{32}>0 \\
                      \psi_{12}<0
                    \end{array}\right.\;\mbox{or}\;
            \left\{\begin{array}{c}
                      \psi_{32}<0 \\
                      0<\psi_{12}\psi_{32}\leqslant9
                    \end{array}\right.\vspace{.2cm}\\
            U_-,\;\;\mbox{for}\;
            \left\{\begin{array}{c}
                      \psi_{32}<0 \\
                      \psi_{12}\psi_{32}\leqslant9
                    \end{array}\right.
          \end{array}\right. \vspace{.2cm}\\
\mbox{Substitution into equation of motion}\\
\mbox{and constraint leads to} \vspace{.2cm}\\
\begin{array}{l}
      \mbox{{\bf a)}}\;\;\omega = -1,\quad\varkappa=-15\zeta_2x^4-10\zeta_1x^2,\quad\dim\(\Omega_3^{\rm sol}\)=2 \vspace{.2cm}\\
      \mbox{{\bf b)}}\;\;\varkappa=0,\quad\psi_{12}\psi_{32}=\dac{27}{4},\quad\dim\(\Omega_3^{\rm sol}\)=1  \vspace{.2cm}\\
      \end{array}
\end{array}
\vrule\hspace{.1cm}
\begin{array}{c}
      \boxed{L_1+L_2} \vspace{1.2cm}\\
      \mbox{necessary conditions:} \vspace{.2cm}\\
      x^2=-\dac{\psi_{12}}{6},\;\psi_{12}<0,\quad y\in\mathbb{R}  \vspace{.2cm}\\
      \mbox{Substitution into equation of motion} \\
      \mbox{and constraint leads to} \vspace{.2cm}\\
      \omega = -1,\;\varkappa=-\dac{5\psi_{12}}{4},\;\dim\(\Omega_2^{\rm sol}\)=1 \vspace{1.5cm}
\end{array}\label{add_case_6_7}}
\hsep
{\bf 8.} $(6+0)=(x,x,x,x,x,x)$
\eq{\mbox{Let us denote}\quad V_{\pm}=\dac{1}{2\psi_{32}}\[-9\pm\sqrt{81-12\psi_{12}\psi_{32}}\],\;\;
W_{\pm}=\dac{1}{6\zeta_2}\[-\zeta_1\pm\sqrt{\zeta_1^2-\dac{4}{5}\varkappa\zeta_2}\]}
\eq{\begin{array}{c}
      L_1+L_2+L_3 \vspace{.5cm}\\
      \mbox{Substitution $H_i\equiv x$ into equation of motion and} \\
      \mbox{constraint leads to} \vspace{.2cm}\\
      \begin{array}{l}
      \mbox{{\bf a)}}\;\;\omega = -1,\quad\varkappa = -5\zeta_3x^6-45\zeta_2x^4-15\zeta_1x^2 \vspace{.2cm}\\
      \mbox{{\bf b)}}\;\;\varkappa=0,\;
      x^2=\left[\begin{array}{l}
            V_+,\;\;\left\{\begin{array}{c}
                      \psi_{32}>0 \\
                      \psi_{12}\psi_{32}\leqslant6
                    \end{array}\right.\;\mbox{or}\;
            \left\{\begin{array}{c}
                      \psi_{32}<0 \\
                      6\leqslant\psi_{12}\psi_{32}\leqslant\dac{27}{4}
                    \end{array}\right. \vspace{.2cm}\\
            V_-,\;\;\psi_{32}<0,\;\psi_{12}\psi_{32}\leqslant\dac{27}{4}
          \end{array}\right.
      \end{array}
      \vspace{1cm}\\
    \end{array}\vrule\hspace{.1cm}
    \begin{array}{c}
      L_1+L_2 \vspace{.2cm}\\
      \mbox{Substitution $H_i\equiv x$ into equation} \\
      \mbox{of motion and constraint leads to} \vspace{.2cm}\\
      \begin{array}{l}
      \mbox{{\bf a)}}\;\;\omega = -1, \\
      x^2=\left[\begin{array}{l}
            W_+,\;\left\{\begin{array}{c}
                      \zeta_1<0 \\
                      \zeta_2>0
                    \end{array}\right.,\;
            \varkappa\leqslant\dac{5\zeta_1^2}{4\zeta_2}\vspace{.2cm}\\
            W_-,\;\zeta_1<0\;\mbox{or}\;
            \left\{\begin{array}{c}
                      \zeta_1>0 \\
                      \zeta_2<0
                    \end{array}\right.
          \end{array}\right.\\
      \mbox{{\bf b)}}\;\;\varkappa=0,\;x^2=-\dac{\psi_{12}}{3},\;\psi_{12}<0
      \end{array}
    \end{array}
\label{add_case_6_8}}

Similar to the (7+1)-dimensional case, one can see that \Lall case is more abundant with solutions. We summarize all solutions obtained in Table
\ref{6.plus.1} and will discuss the differences and familiarities in detail in the appropriate section.

\begin{table}[!h]
\begin{center}
\caption{Summary of (6+1)-dimensional solutions in \Lall and EGB.}
\label{6.plus.1}
  \begin{tabular}{|c|c|c|c|c|}
    \hline
    Splitting & nonvacuum \Lall & vacuum \Lall & nonvacuum EGB & vacuum EGB  \\
    \hline
    $(2-2+1+1)$ (\ref{add_case_6_1}) & yes & \multicolumn{3}{c|}{no} \\
    \hline

    $(2+2+2)$ (\ref{add_case_6_2}) & yes & \multicolumn{3}{c|}{no} \\
    \hline
    $(3+2+1)$ (\ref{add_case_6_3}) & \multicolumn{3}{c|}{yes} & no \\
    \hline
    $(4+1+1)$ (\ref{add_case_6_4}) & yes & \multicolumn{3}{c|}{no} \\
    \hline
    $(3+3)$ (\ref{add_case_6_5}) & \multicolumn{4}{c|}{yes} \\
    \hline
    $(4+2)$ (\ref{add_case_6_6}) & \multicolumn{4}{c|}{yes} \\
    \hline
    $(5+1)$ (\ref{add_case_6_7}) & \multicolumn{3}{c|}{yes} & no \\
    \hline
    $(6+0)$ (\ref{add_case_6_8}) & \multicolumn{4}{c|}{yes} \\
    \hline
  \end{tabular}
\end{center}
\end{table}

\section{Spaces with three-dimensional isotropic subspace}

Now we want to turn our attention to the special case with three-dimensional isotropic subspace. Despite the fact that in the previous section we gave
complete description of all possible cases,
due to importance of the case with three-dimensional isotropic subspace -- it could give a rise to successful dynamical compactification scenario where
three dimensions expand isotropically while the
remaining are contracting, and they can contract anisotropically -- we decided to devote a separate section for this comparison.

In (7+1)-dimensional \Lall case we can have above mentioned scenario as a special case of (\ref{case_7_1}) when one of the $\{y,\,z,\,u\}$ is equal to $x >
0$
(turning it into $(3-2+1+1)$) and the remaining two are negative.
For $(3-3+1)$ case (\ref{case_7_2}) we require $x > 0$ with $y < 0$ and with no additional constraints retrieve dynamical compactification of all but three
space directions.
Next case -- $(3+3+1)$ (\ref{case_7_3}) -- once again could be easily compactified -- for nonvacuum solution we require $x > 0$ with $z < 0$ (we can always
do this since $z\in\mathbb{R}$); we also need
$y < 0$ which could be achieved with $\psi_{32} > 0$ (see (\ref{case_7_3})) and so $\psi_{12} \psi_{32} \leqslant 1$. For vacuum solution one can see that
both branches for $\xi$ satisfy dynamical
compactification as long as $z < 0$.

The case $(3+2+2)$ (\ref{case_7_4}) is more interesting -- indeed, for nonvacuum case we need $B_1(x) \geqslant 0$ while $y,\,z < 0$ for $x>0$; analytical
solution of this inequality is next to
impossible to obtain so we enumerate possibilities over a reasonable range of $\psi_{12}$, $\psi_{32}$ and $x>0$. Our analysis suggests that there is a
upper bound on $x$:
$x \leqslant 6.5$\footnote{Keep (\ref{add.norm}) in mind for physical value of $x$.} and
allowed regions for $\psi_{12}$ and $\psi_{32}$ depends on $x$. One can simply verify that vacuum \Lall solution is also fine. Nonvacuum EGB solution is
fine as long as $y,\,z<0$ -- constraints
on both could be easily derived from (\ref{case_7_4}).

Finally, $(4+3)$ case (\ref{case_7_6}), and now we require $x<0$ while $y>0$. Similarly to the previous case we numerically investigated allowed regions on
$\{\psi_{32},\,\psi_{12},\,x\}$ space,
which lead to compactification. We found that for any $x<0$ there exists region on $\{\psi_{32},\,\psi_{12}\}$ plane which lead to $y > 0$, so that the
limit similar to the previous case does not exist
in this one. To support our claim we rewrite (\ref{case_7_6}) for $\alpha$ and $\beta$ (keep in mind (\ref{add.relation})):

\begin{equation}
\begin{array}{l}
\alpha = - \dac{9y^3 + 23y^2x + 27yx^2 + 12x^3}{64xy^4 + 204x^2y^3 + 264x^3y^2 + 144x^4y + 25x^5}, \vspace{.3cm}\\
\beta=\dac{1}{24} \dac{5x^4 + 12x^3y + 11x^2y^2 + 6xy^3 + y^4}{yx^3 (6x^4 + 36x^3y + 66x^2y^3+51xy^3 + 16y^4)},\quad x,\,y\in\mathbb{R}.
\end{array} \label{case_7_6_1}
\end{equation}

As for EGB case, one can easily see from (\ref{case_7_6}) that both vacuum and $\Lambda$-term case always
have successful dynamical compactification: for nonvacuum with $x<0$ we can choose branch $y=-3x+\sqrt{6x^2-\psi_{12}}$ to ensure $y > 0$ while for vacuum
both $\xi$ give opposite signs to $x$ and $y$.

We summarize all possible solutions in Table \ref{7.plus.1.comp}. One can clearly see that the situation with solutions which allow ``good''
compactification follows the general trend -- we have \Lall case more abundant on the solutions; as of the comparison of $\Lambda$-term solutions with
vacuum, the former are more abundant then the latter.

\begin{table}[h]
\begin{center}
\caption{\centering Summary of (7+1)-dimensional solutions which could lead to successful dynamical compactification in \Lall and EGB.}
\label{7.plus.1.comp}
  \begin{tabular}{|c|c|c|c|c|}
    \hline
    Splitting & nonvacuum \Lall & vacuum \Lall & nonvacuum EGB & vacuum EGB  \\
    \hline
    $(3-2+1+1)$ (\ref{case_7_1}) & $\begin{array}{c}
                                     \mbox{special choice, say,} \\
                                     y=x>0,\; z,\,u < 0
                                   \end{array}$ & \multicolumn{3}{c|}{no} \\
    \hline
    $(3-3+1)$ (\ref{case_7_2}) & \multicolumn{3}{c|}{$y < 0$} & no \\
    \hline
    $(3+3+1)$ (\ref{case_7_3}) & $z<0$, $\psi_{32} > 0$, $\psi_{12} \psi_{32} \leqslant 1$ & $z<0$ & \multicolumn{2}{c|}{no} \\
    \hline
    $(3+2+2)$ (\ref{case_7_4}) & $x \lesssim 6.5$ (see test for details) & yes & constraints from $y<0$ & no \\
    \hline
    $(4+3)$ (\ref{case_7_6}) & \multicolumn{4}{c|}{yes (see text for details)} \\
    \hline
  \end{tabular}
\end{center}
\end{table}

Now let us perform the same analysis for (6+1)-dimensional solutions. The case with $(2-2+1+1)$ splitting has only nonvacuum \Lall case
(\ref{add_case_6_1}) and it could be
brought to $(3-2+1)$ with special choice of $\{y,\,z\}=x > 0$, while the remaining Hubble parameter is negative.
Nonvacuum \Lall $(3+2+1)$ case (\ref{add_case_6_3}) requires additional constraint coming from $y < 0$ requirement plus we require $z<0$; vacuum \Lall and
nonvacuum EGB
cases satisfied by default: the former of them has the only solution which already has opposite signs for $x$ and $y$ (plus $z<0$) while for the latter we
just need $z<0$.
Final case to consider is $(3+3)$ (\ref{add_case_6_5}) and there are four subcases: nonvacuum \Lall -- using enumeration technique in reasonable range for
$x>0$, $\psi_{12}$ and $\psi_{32}$
we proved that for any $x>0$ there exist area in $\{\psi_{12},\,\psi_{32}\}$ space which corresponds to $y<0$; additionally we can, similarly to $(4+3)$
case, rewrite (\ref{add_case_6_5}) for
$\alpha$ and $\beta$ (again, keep in mind (\ref{add.relation})):

\begin{equation}
\begin{array}{l}
\alpha = - \dac{3x^2 + 4xy + 3y^2}{16x^3y + 28x^2y^2 + 16xy^3},~\beta=\dac{1}{24} \dac{x^4 + 4x^3y + 5x^2y^2 + 4xy^3 + y^4}{x^3y^3(4x^2 + 7xy +
4y^2)},~x,\,y\in\mathbb{R}.
\end{array} \label{case_6_5_1}
\end{equation}

Vacuum \Lall case -- from $P_2(x,\,y) = 0$ with use of $y=\xi x$ substitution we
get $\xi\in \[-1.5;\,0\)$ so that we also have compactification; nonvacuum EGB: $y=-2x-\sqrt{3x^2-\psi_{12}}$ branch always gives us $y<0$ for $x>0$; and
finally vacuum EGB case also does
possess successful compactification since $y=\xi x$ and $\xi = - (3\pm\sqrt{5}/2 < 0)$ both. So that all $(3+3)$ cases have dynamical compactification.
Again we summarize all possible solutions in Table \ref{6.plus.1.comp}.

\begin{table}[h]
\begin{center}
\caption{\centering Summary of (6+1)-dimensional solutions which could lead to successful dynamical compactification in \Lall and EGB.}
\label{6.plus.1.comp}
  \begin{tabular}{|c|c|c|c|c|}
    \hline
    Splitting & nonvacuum \Lall & vacuum \Lall & nonvacuum EGB & vacuum EGB  \\
    \hline
    $(3-2+1)$ (\ref{add_case_6_1}) & special choice, say, $y=x>0$, $z < 0$ & \multicolumn{3}{c|}{no} \\
    \hline
    $(3+2+1)$ (\ref{add_case_6_3}) & constraints from $y<0$ + $z<0$ & \multicolumn{2}{c|}{$z<0$}  & no \\
    \hline
    $(3+3)$ (\ref{add_case_6_5}) & \multicolumn{4}{c|}{yes (see text for details)} \\
    \hline
  \end{tabular}
\end{center}
\end{table}

\section{Conclusions}

In this paper we studied higher-dimensional exponential solutions for EGB and Lovelock gravity with third correction. As we mentioned in the Introduction,
both cases -- EGB and mode general Lovelock
gravity -- are eligible and well-motivated, and they demonstrate different properties of the solutions. We developed a scheme which allows one to find all
possible exponential solutions in any number
of dimensions and with all possible Lovelock terms. The scheme was tested on (6+1)- and (7+1)-dimensional space-times and all possible solutions for them
were found -- they are summarized in
Tables \ref{7.plus.1} and \ref{6.plus.1}. We can see that these solutions are more abundant in the presence of all three Lovelock terms
as well as in the presence of non-zero cosmological constant in the action.  In the most general case in (7+1)-dimensions we have 10 possible
splitting of initial 7-dim space into different isotropic subspaces (which number does not exceed 5), a vacuum general third order Lovelock
theory has 7 possibilities (with no more than 3 isotropic subspaces), non-vacuum EGB theory has 6 cases,  and a vacuum EGB - only 3 of them
(the latter theory has solutions only with two different subspaces).  The same tendency appears in the (6+1)-dimensional case with 8 solutions
in the general non-vacuum case and 3 solutions in the vacuum EGB theory.

As a special case we separate the solutions which allow three-dimensional isotropic subspace with different signs for Hubble parameters corresponding to
this subspace and all remaining directions --
this requirement ensures that three-dimensional subspace will expand (given positive sign for the corresponding Hubble parameter) while in the remaining
directions Universe will contract. In turn,
this allows us to consider this situation as successful example of dynamical compactification -- indeed, we have three dimensions expanding and the
remaining contracting -- though, they could
contract anisotropically. We summarized our findings of the solutions of this type in Tables \ref{7.plus.1.comp} and \ref{6.plus.1.comp}.
Similar to general solutions, the more terms in the action of the theory we consider, the more possibilities for compactification we have.

This finalize first part of our study of exponential solutions in Lovelock gravity. By now we completed investigation of exponential solutions in EGB gravity, described both non- and constant-volume
solutions. In this paper we generalized non-constant-volume solutions on the case of Lovelock gravity. By comparing solutions and their abundance for EGB and \Lall one clearly see that more general 
Lovelock case has more different solutions, which is partially due to increased number of degrees of freedom (through increased number of couplings), and with growth of the number of dimensions the
highest possible Lovelock order is also increasing, resulting in growing number of possible solutions. This makes the same solution with three-dimensional isotropic subspace more ``probable'' to
appear in case with more Lovelock corrections taken into account, which, in turn, could be used as kind of anthropic argumentation for consideration of Lovelock and other higher-order gravity models.

\begin{acknowledgments}
The work of A.T. is supported by RFBR grant 14-02-00894 and partially supported by the Russian Government Program of Competitive Growth of Kazan Federal University.
S.A.P. is supported by FONDECYT via grant No. 3130599.
\end{acknowledgments}

\end{document}